\documentclass[journal,twocolumn,twoside,10pt]{IEEEtran}
\usepackage[left=1.7cm, right=1.7cm, top=2cm, bottom=2cm]{geometry}

\usepackage[utf8]{inputenc}
\usepackage{microtype,multibib}
\usepackage{graphicx}
\usepackage{booktabs}
\usepackage{amsmath}
\usepackage{amsfonts}
\usepackage{amsthm}
\usepackage{commath,todonotes}
\usepackage{hyperref}
\usepackage{caption}
\usepackage{subcaption}
\usepackage{cite}

\newcommand{\RR}{\mathbb{R}}

\newcommand{\NN}{\mathbb{N}}

\newcommand{\CC}{\mathbb{C}}
\renewcommand{\tilde}{\widetilde}

\usepackage{xcolor}
\newcommand{\reviewed}[1]{{\color[rgb]{0,0,0}{#1}}}

\title{Time-Frequency Phase Retrieval for Audio --- \\The Effect of Transform Parameters}
\author{Andr\'es Marafioti, Nicki Holighaus, and Piotr Majdak
\renewcommand\footnotemark{}

\thanks{Manuscript received on November 2020; Revised in April 2021.}%
\thanks{The authors are with the Acoustics Research Institute, Austrian Academy of Sciences, Wohllebengasse 12--14, 1040 Vienna, Austria.}
\thanks{Accompanying web page: 
\texttt{\url{https://github.com/andimarafioti/phaseRetrievalEvaluation}}. \newline We thank Nathana\"{e}l Perraudin for the fruitful discussions we had over the years about the importance of the quality of phase retrieval in more complex systems. 
This work has been supported by Austrian Science Fund (FWF) project MERLIN (Modern methods for the restoration of lost information in digital signals;I 3067-N30).}}
\makeatletter
\def\ps@IEEEtitlepagestyle{%
  \def\@oddfoot{\mycopyrightnotice}%
  \def\@evenfoot{}%
}
\def\mycopyrightnotice{%
  {\footnotesize \begin{minipage}{\textwidth} \copyright 2021 IEEE.  Personal use of this material is permitted.  Permission from IEEE must be obtained for all other uses, in any current or future media, including reprinting/republishing this material for advertising or promotional purposes, creating new collective works, for resale or redistribution to servers or lists, or reuse of any copyrighted component of this work in other works.\end{minipage}}
}
\makeatother

\begin{document}

\maketitle
\begin{abstract}

In audio processing applications, phase retrieval (PR) is often performed from the \reviewed{magnitude of} short-time Fourier transform (STFT) coefficients. Although PR performance has been observed to depend on the considered STFT parameters and audio data, the extent of this dependence has not been systematically evaluated yet. To address this, we studied the performance of three PR algorithms for various types of audio content and various STFT parameters such as redundancy, time-frequency ratio, and the type of window. The quality of PR was studied in terms of objective difference grade and signal-to-noise ratio of the STFT magnitude, to provide auditory- and signal-based quality assessments. Our results show that PR quality improved with increasing redundancy, with a strong relevance of the time-frequency ratio. The effect of the audio content was smaller but still observable. \reviewed{The effect of the window was only significant for one of the PR algorithms.} Interestingly, for \reviewed{a good} PR quality, each of the three algorithms required a different set of parameters, demonstrating the relevance of individual parameter sets for a fair comparison across PR algorithms. Based on these results, we \reviewed{developed guidelines} for \reviewed{optimizing} STFT parameters \reviewed{for a given application}.

\end{abstract}

\section{Introduction}

\reviewed{
Phase is a crucial component of audio signals and affects how humans perceive sounds~\cite{laitinen2013sensitivity} and speech~\cite{liu1997effects, paliwal2005usefulness}. When processing audio, a signal is often represented in the complex-valued short-time Fourier transform~(STFT) domain~\cite{oppenheim1981importance,Allen1977,rawe90}), although many audio applications focus on processing the STFT magnitude~\cite{vincent2018audio,Chowdhury2019,pepino2018sourcesep, ghose2020Enabling}. In order to synthesize the targeted time-domain signal, they estimate the STFT phase from the processed STFT magnitudes by performing phase retrieval (PR)~\cite{magron2018reducing, Liu2020Unified}. The necessity of PR also arises in the generation of a signal described only by STFT magnitude~\cite{marafioti2020gacela, engel2019gansynth}. PR algorithms have been used successfully in the field of audio~\cite{gerkmann2015phase,mowlaee2017single,magron2015phaserecovery}, including specific applications such as audio inpainting \cite{marafioti2019context, marafioti2019audio}, but many applications exist beyond the audio domain, e.g., in X-ray crystallography~\cite{harrison1993phase, miao2008extending} and imaging~\cite{fogel2016phase, shechtman2015phase}.  

The input to PR algorithms is usually given by phaseless transform coefficients with respect to some dictionary. The classic problem of Fourier-based PR~\cite{Bendory2017} has been extended to deal with various time-frequency (TF) representations, most notably the STFT, the best understood and most widely used TF representation in the field of audio processing. However, different STFT parameters may yield different PR results. From the mathematical perspective, PR is a difficult inverse problem and various conditions ensuring its feasibility have been derived, e.g., ~\cite{balan2006signal,nawab1983signal,bojarovska2016phase,jaganathan2016stft,li2017phase,alaifari2019ill, waldspurger2018phase}, yielding conditions on the window, transform parameters, or limitations in the processed input. These conditions, while theoretically correct, can be impractical in actual applications. To find practical conditions, we evaluated the performance of various PR algorithms under systematic variation of STFT parameters and audio type.

Phase\footnote{From now on, we use \textit{phase} when referring to the STFT phase.} retrieval for audio signals reached its first milestone in 1984, when the Griffin-Lim algorithm \reviewed{(GLA)} was introduced~\cite{griffin1984signal}. It is iterative and computationally intensive in each iteration and unsuitable for real-time applications. Further  improvements with respect to quality~\cite{masuyama2020deepGriffin, masuyama2019deep, Masuyama2020ICASSP} and computation time~\cite{Masuyama2019griffinLim,le2010fast} yielded algorithms such as the fast Griffin-Lim algorithm (FGLA)~\cite{Perraudin2013griffin} \reviewed{and real-time iterative spectrogram inversion~\cite{Zhu2006}}. Nowadays, GLA is a widely used iterative PR algorithm, e.g.,~\cite{mimilakis2018monaural, vasquez2019melnet}. 

Alternative, non-iterative algorithms such as single-pass spectrogram inversion (SPSI)~\cite{beauregard2015single}, phase unwrapping~\cite{magron2015phase}, and phase gradient heap-integration (PGHI)~\cite{ltfatnote040}, have been proposed. SPSI assumes a sinusoidal model with linear phase progression and phase locking~\cite{laroche1999improved} to the closest spectral peak. It is fast and directly suitable for real-time usage, but it relies on the assumption that the signal consists of slowly varying sinusoidal components. PGHI is efficient and equally suitable for real-time processing. Even though PGHI does not rely on any signal assumptions, it is based on the phase-magnitude relations~\cite{portnoff1979magnitude}, a property of the continuous STFT, which when used in a discrete realm may introduce inaccuracies governed by the parameters of the discrete STFT~\cite{marafioti2019adversarial}.

All these algorithms are applied to magnitude STFT coefficients which can be obtained with different parameters. Unfortunately, besides general introductions to phase-aware processing~\cite{mowlaee2016advances, gerkmann2015phase}, there are only a few hints towards which are `good' STFT parameters for PR algorithms. For example, a large number of frequency channels seems to be beneficial in music applications~\cite{engel2019gansynth}. For a fixed number of frequency channels, the window function and redundancy seems to affect the PR quality~\cite{ltfatnote043}, with larger redundancy improving the PR quality~\cite{holighaus2019char,holighaus2019non}. Still, a systematic investigation of the transform parameters affecting the PR quality of audio seems to be missing.

In this article, we first revisit relevant properties of the discrete STFT \cite{po76,auger2012phase,gr01} in the context of PR. We then systematically evaluate three PR algorithms: PGHI, FGLA, and SPSI. We consider five redundancies of the STFT between $2$ and $32$, many window sizes ranging from $32$ to $61440$ samples, and four types of windows: Gaussian, Blackman, Hann, and Bartlett. In addition, we evaluate the PR performance in a simple setting with processed magnitude spectrograms. We also consider the performance of PR for various types of audio signals. Finally, we describe guidelines for obtaining good STFT parameters for a given PR algorithm. The code used to perform our experiments is available at \url{https://github.com/andimarafioti/phaseRetrievalEvaluation}.
}
\section{The discrete short-time Fourier transform}\label{sec:discSTFT}

We consider finite signals $s\in\CC^L$ and indices in the signal domain are to be understood modulo $L$.
The STFT of $s$, with the \emph{analysis window} $g\in\RR^L$, time step $a\in \NN$ and $M\in\NN$ frequency channels is given by 
\begin{equation}\label{eq:STFT_final}
\begin{split}
   \operatorname{S}_g(s)[m,n] & = \sum_{l=0}^{L-1} s[l]g[l-na]e^{-2\pi i ml/M}\\
   & = \left|\operatorname{S}_g(s)[m,n]\right|e^{i\mathrm{\phi}_g(s)[m,n]}, 
\end{split}
\end{equation}
for
$n\in \{0,\ldots,L/a-1\}$ 
and $m\in\{0,\ldots,M-1\}$. If $s$ and $g$ are real-valued, the STFT is conjugate symmetric in $m$ and it is sufficient to store the first $M_{\RR} = \lfloor (M/2)+1\rfloor$ channels. Note that $\mathrm{\phi}_g(s)$ refers to the TF phase, which we refer to simply by 
\textit{phase} throughout this document. Accordingly, TF PR is concerned with estimating the phase, or equivalently $\operatorname{S}_g(s)$, from the \emph{magnitude} $\left|\operatorname{S}_g(s)\right|$.


\subsection{Properties of the STFT}
\label{subsec:resolution_stft}

\begin{figure}[t!]
\centerline{\includegraphics[width=1\columnwidth]{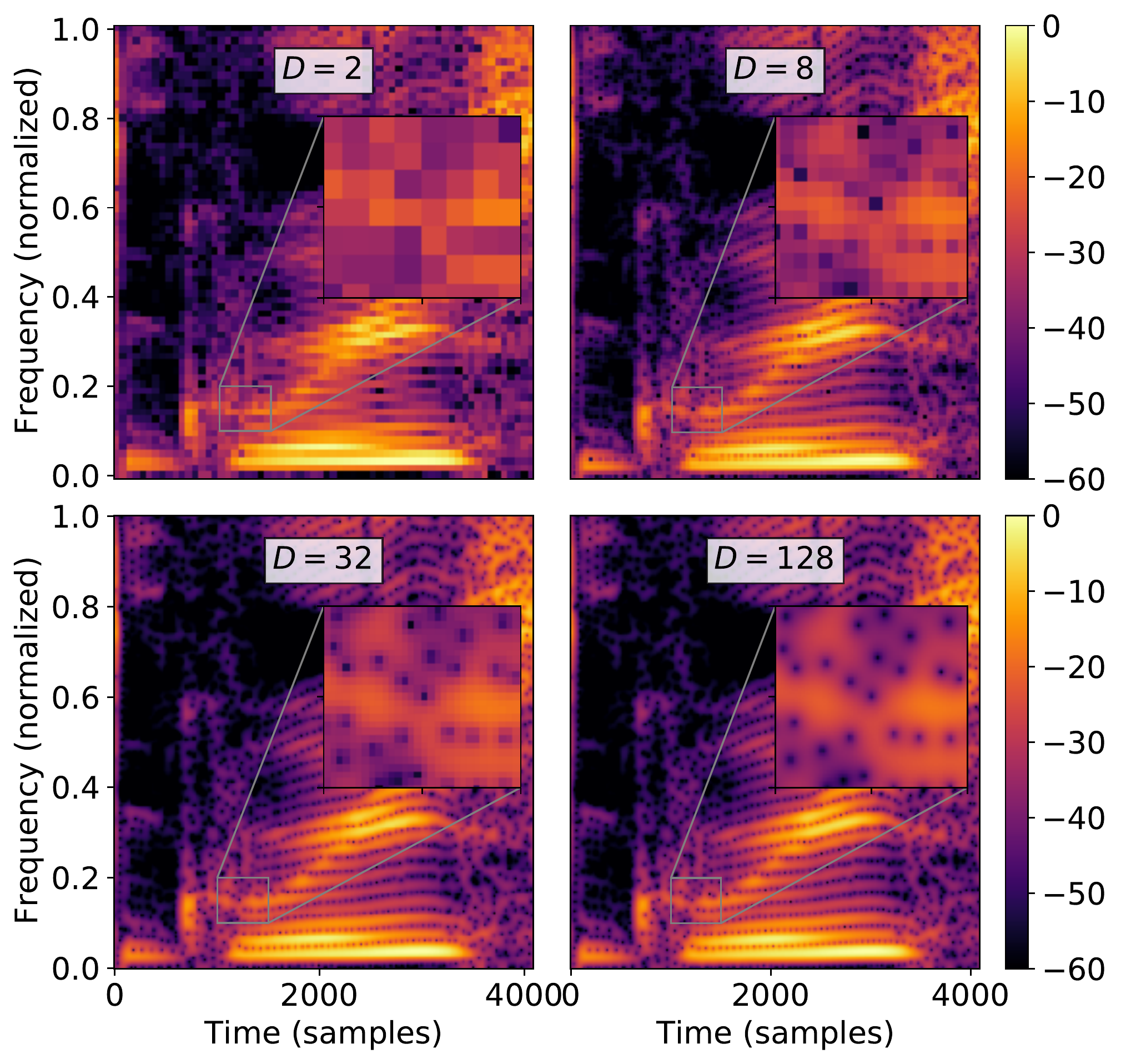}}
\caption{Exemplary spectrograms calculated for various redundancies $D$. Calculations were done with the Gaussian window and TF ratio of $\lambda=8$.}
\label{fig:spec_sametfr}
\end{figure}

Depending on the choice of transform parameters $a$, $M$, and the window $g$, the discrete STFT encodes time and frequency information with different properties. The \emph{full} STFT, i.e., with $a=1$ and $M=L$, is a slowly varying function, owing to significant overlap between both the time range covered by adjacent time positions and the frequency range covered by adjacent frequency channels. When increasing the time step $a$ over $1$, the \textit{time resolution} of the STFT decreases. Similarly, when decreasing the number of channels $M$ below $L$, the \textit{frequency resolution} of the STFT decreases. 
Jointly, time and frequency resolution can be likened to the pixel resolution in digital imaging. 
This joint resolution is characterized by the \textit{redundancy (D)} of the STFT, $D=M/a$.  Figure~\ref{fig:spec_sametfr} shows examples of STFT magnitudes calculated with the same window $g$, for various redundancies $D$. Especially at redundancy $D=2$, it can be seen that some characteristic features of the STFT magnitude are obscured.

Further, the window $g$ and it's Fourier transform $\widehat{g}$ control the inherent TF uncertainty~\cite{gr01} of the STFT, independently of $a$ and $M$. Namely, every window function $g$ has a certain shape in time and $\widehat{g}$ in frequency, which determine how spectro-temporal signal components are \emph{smeared} in the STFT magnitude. The shape of a window is usually characterized by its width. In the classic uncertainty principle, a window's time and frequency width are defined as the standard deviation of $g$ and of $\widehat{g}$, respectively. This notion is reasonable for any smooth, roughly bell-shaped window. 

\begin{figure}[t!]
\centerline{\includegraphics[width=0.95\columnwidth]{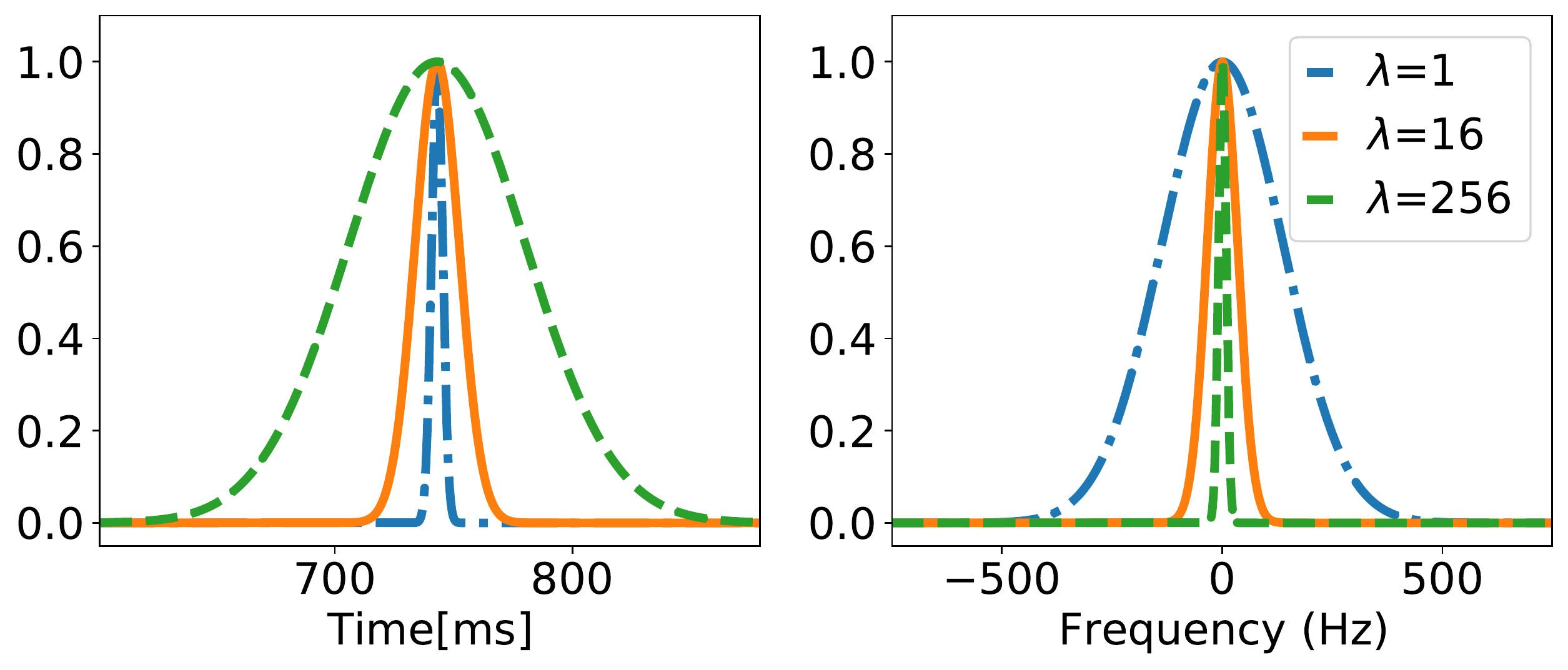}}
\caption{Examples of Gaussian windows in the time domain (left) and magnitude of their Fourier transform (right) for various TF ratios $\lambda$. The length of the windows was the same, i.e. $L=65536$ samples, in all examples.
}
\label{fig:resolution_samered}
\end{figure}

\begin{figure}[th!]
\centerline{\includegraphics[width=\columnwidth]{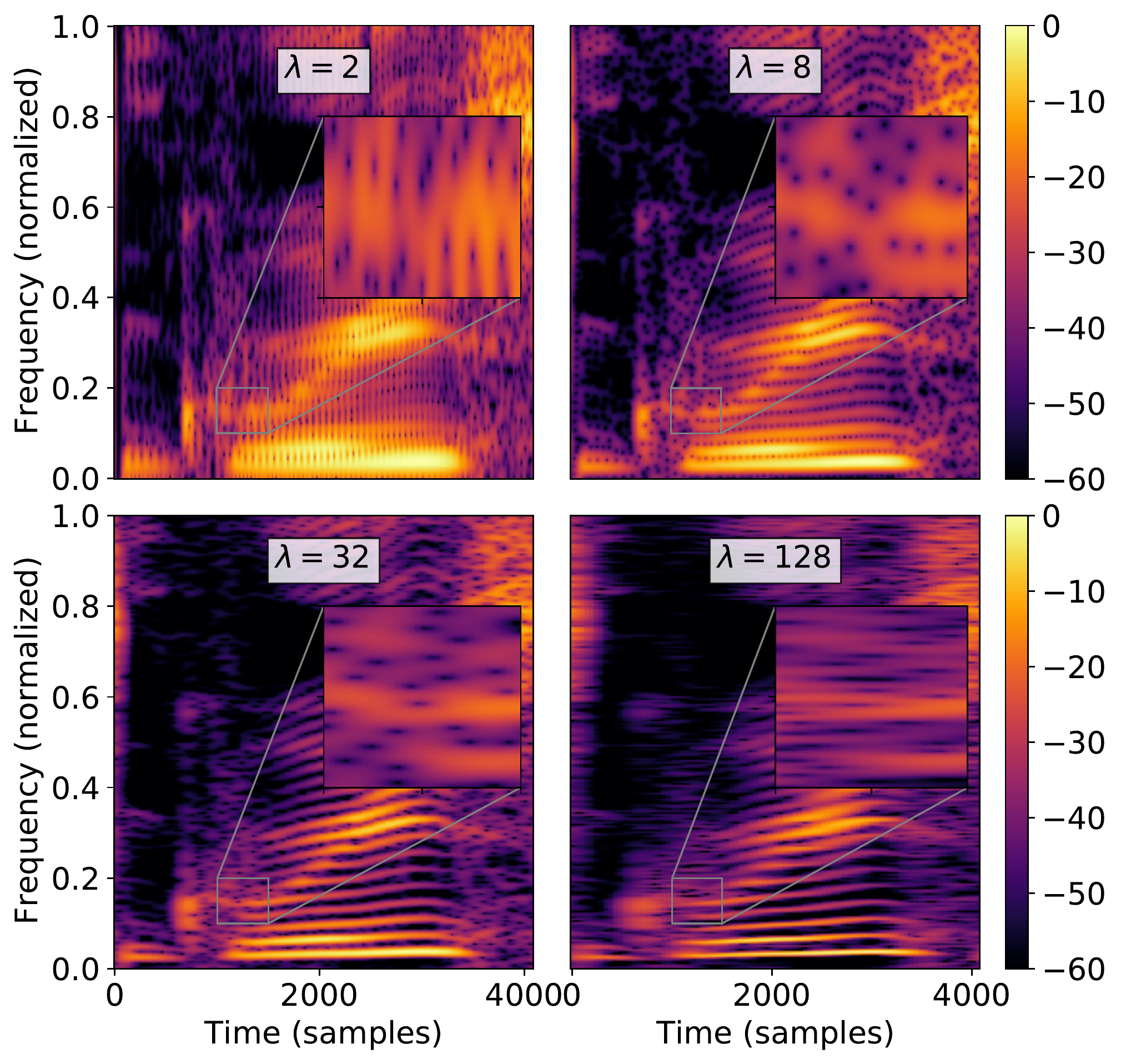}}
\caption{Exemplary spectrograms calculated for various TF ratios $\lambda$ of the Gaussian window. The same redundancy of $D=128$ was used in all examples.}
\label{fig:spec_samered}
\end{figure}

The classic example of a bell-shaped window is the Gaussian window. The Gaussian window minimizes the product of time and frequency width and its Fourier transform is a Gaussian as well. The \emph{discrete, periodized} Gaussian, simply referred to as \emph{Gaussian} in this study, is defined as:
\begin{equation}\label{eq:Gaussian_Lambda}
    \large
      g_\lambda[l] := \sum_{k=-\infty}^{\infty} e^{-\frac{\pi (l-kL)^2}{\xi_s\lambda}}, \quad \lambda, \xi_s\in\RR^+,
\end{equation}
where $\xi_s$ is the assumed sampling rate (in Hz). It inherits from its continuous counterpart the property that its DFT $\widehat{g_\lambda}$ is again a Gaussian. 
The parameter $\lambda$ defines jointly the width of $g_\lambda$ and $\widehat{g_\lambda}$, \reviewed{as illustrated in Figure \ref{fig:resolution_samered} by the inverse relation between the width of $g_\lambda$ and $\widehat{g_\lambda}$.}
Precisely, the width of $g_\lambda$ (measured in samples) is $\lambda$ times as large as the width of $\widehat{g_\lambda}$ (measured in Hz). This is why $\lambda$ can be referred to as the \emph{TF ratio} of a Gaussian window. The effect of $\lambda$ on the STFT is shown in Figure~\ref{fig:spec_samered}, for different STFTs at the same redundancy $D$.

The Gaussian window is available in public libraries such as Scipy~\cite{2020SciPy-NMeth} and LTFAT~\cite{ltfatnote030}. However, it is not the most commonly used window function. Therefore, libraries specializing in a particular field often do not implement it, e.g., PyTorch~\cite{pytorch} 
for machine learning. Instead, 
it is more common to compute the STFT 
using windows with short support, such as the \reviewed{Hann, Hamming or rectangular windows}. For those windows $g$ there is no exact equivalent to the TF ratio $\lambda$. Instead, one can determine \reviewed{their equivalent $\lambda$} through comparison to the Gaussian window. \reviewed{Precisely, given a window $g$, we find $\lambda$ as $\mathop{\operatorname{argmin}}_\lambda \|g - g_\lambda\|$, assuming that $g$ is peak-normalized, i.e., $\max |g[l]|=1$. Alternatively, $g$ can be fit to a given $\lambda$ by adjusting the length of $g$ to minimize the norm distance to $g_\lambda$.}
%

In conclusion, the overall numerical properties of the discrete STFT depend on the \reviewed{joint choice of the parameters $g_\lambda$, governing its uncertainty, and $a$ and $M$, controlling its resolution}. The most favorable properties can be achieved when the uncertainty is matched to the resolution, \cite{best03,faulhuber2017optimal}. With the definitions in Eqs. \eqref{eq:STFT_final} and \eqref{eq:Gaussian_Lambda}, uncertainty and resolution are matched if and only if 
\begin{equation}\label{eq:Lambda}
    \lambda = aM/\xi_s.
\end{equation}
\reviewed{In this case, the STFT samples lie on a grid on which the ratio between time- and frequency-steps coincides with $\lambda$, leading to an optimally uniform covering of time-frequency space.}
In all our experiments, $\lambda$, $a$ and $M$ are linked in this fashion. 

\subsection{Inverse STFT for signal synthesis}

For any \emph{synthesis window} $\tilde{g}\in\RR^L$, the inverse STFT of $S\in\CC^{M\times N}$ with respect $\tilde{g}$ is given by 
\begin{equation}\label{eq:iSTFT_discrete}
\begin{split}
   \tilde{s}[l] = \sum_{n=0}^{N-1} \sum_{m=0}^{M-1} \operatorname{S}[m,n]\tilde{g}[l-na]e^{2\pi iml/M}, 
\end{split}
\end{equation}
for $l\in\{0,\ldots,L-1\}$. If a $\tilde{g}$ exists, such that $\tilde{s} = s$ for all $s\in\CC^L$, and with $\operatorname{S} = \operatorname{S}_g(s)$, then the STFT $\operatorname{S}_g$ is invertible, i.e., it forms a frame in the sense of~\cite{rawe90,st98-1,janssen1997continuous} and $\tilde{g}$ is a \emph{dual window} for $g$. Generally, in order to obtain an invertible STFT, a redundancy equal to or larger than one, $D=M/a \geq 1$ is required\footnote{In contrast to common practice, the number of channels $M$ may be smaller than the number of nonzero samples in $g$.}.

\begin{figure}[ht!]
\centerline{\includegraphics[width=0.7\columnwidth]{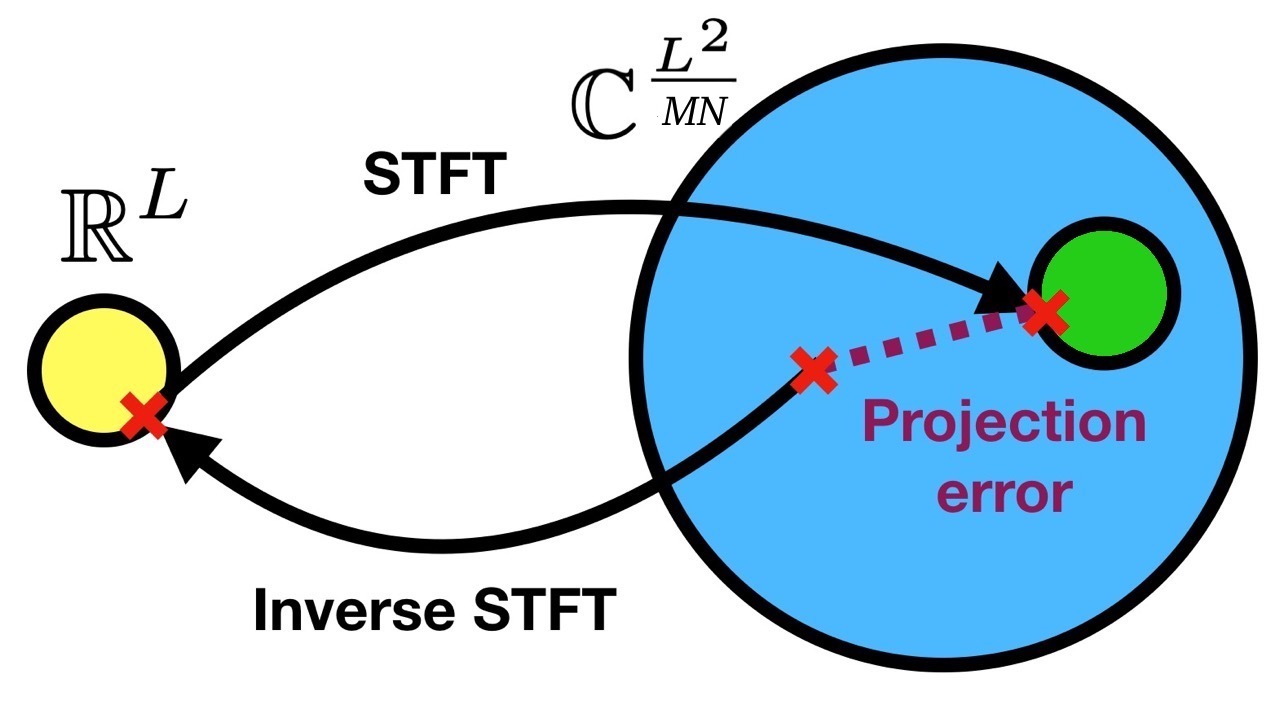}}
\caption{Blue circle: Set of all possible TF coefficients. Yellow circle: set of time-domain signals. Green circle: set of consistent STFT coefficients. An inverse STFT done on a point from the blue set yields a point from the yellow set. An STFT from this point yields a point from the green circle, introducing a projection error. An inverse STFT done on a point from the green circle yields a point from the yellow set, which after a subsequential STFT is remapped to the original point in the green set without any projection errors.}
\label{fig:consistency}
\end{figure}

For redundancies $D>1$, the STFT is overcomplete (or redundant), and $\operatorname{S}_g$ maps into a strict subspace of $\CC^{M\times N}$. In other words, not every matrix $\operatorname{S}\in \CC^{M\times N}$ represents an STFT. We call $\operatorname{S}$ \emph{consistent} if there is a signal $s$, such that $\operatorname{S}=\operatorname{S}_g(s)$ and \emph{inconsistent} otherwise. \emph{Implicitly}, the inverse STFT operation applied to $\operatorname{S}$ projects onto the image of $\operatorname{S}_g$ before synthesis, as visualized in Fig.~\ref{fig:consistency}. In practice, this means that the inverse STFT, applied to inconsistent coefficients  $\operatorname{S}$ produces a signal $\tilde{s}$ with $\operatorname{S_g(\tilde{s})}\neq\operatorname{S}$. Applied to consistent coefficients $\operatorname{S}$, we instead obtain $\operatorname{S} = \operatorname{S}_g(\operatorname{S}_g^{-1}(\operatorname{S}))$. Thus, in the setting of PR, synthesis from a given spectrogram $\left|\operatorname{S}_g(s)\right|$ with a mismatched phase estimate $\mathrm{\phi}$ will often lead to a poor reconstruction. In some cases, we may also use \emph{(in)consistent} to describe magnitude coefficients $\operatorname{M}\in(\RR^+_0)^{M\times N}$that do not correspond to the STFT of \emph{any} signal, i.e., $\operatorname{M} \neq |\operatorname{S}_g(s)|$ for all $s\in\CC^L$.

\section{General methods}

\subsection{Datasets}\label{subsec:data}

The phase of different types of audio signals may have completely different characteristics, thus, we considered three types of signals for the evaluation \reviewed{attempting to cover a wide range of audio characteristics}. First, we included speech signals to the evaluation, because it is a widely used class of signals, it consists not only of harmonic components, but also transients and stochastic segments such as fricatives. 
Second, we considered \reviewed{piano music synthesized from MIDI because it represents a class of simple polyphonic sounds resembling a linear combination of sine waves, without any ambient recording noise}. Third, we considered actual music recordings, which included the natural variations from the musicians and ambient noise from the recording setup.

\textbf{1) Speech.}
For speech, we used LJ Speech \cite{ljspeech17}, which is a public-domain English speech dataset consisting of 13,100 short audio clips of a single speaker reading passages from seven non-fiction books. All clips \reviewed{have a sampling rate of $22050$ Hz and} vary in length from 1 to 10 seconds and have a total length of approximately 24 hours.

\textbf{2) Midi synthesized music.}
We used the Lakh MIDI dataset~\cite{raffelthesis}, a collection of 176,581 unique simple piano MIDI files, to synthesize piano audio signals. This MIDI set was created with the goal of facilitating large-scale music information retrieval, both symbolic (using the MIDI files alone) and audio content-based (using information extracted from the MIDI files as annotations for the matched audio files). The audio files were synthesized from MIDI data using pretty\_midi~\cite{prettymidi}, specifically its fluidsynth API. We generated just one instrument \reviewed{with a sampling rate of $22050$ Hz and} and set it to the piano program 1. 

\textbf{3) Music.}
We segmented the `small' dataset of the free music archive (FMA,~\cite{fma2017}) by genre and used the genre `electronic'. This was done to reduce the variability in the music structure in our evaluations. FMA is an open and easily accessible dataset, usually used for evaluating tasks in musical information retrieval. The \reviewed{subset we used is comprised 1,000 $30$-s segments of songs sampled at $44.1$~kHz.}

\reviewed{The audio material was processed at a sampling rate of $22050$ Hz after resampling to that rate if necessary}. Our experiments use the Large Time-Frequency Analysis Toolbox (LTFAT, \cite{ltfatnote030}).  Hence,  \reviewed{since LTFAT's STFT requires $a, M$ to be divisors of the length of the signal $L$, we chose the first $L=122880$ samples of a randomly selected $128$ signals for each subset, resulting in approximately $5.6$ seconds per signal}. For the experiments with varying number of channels $M$, after reconstruction a portion of signal of length $M$ was removed at the beginning and the end of the signals to avoid issues introduced by the circularity of the considered STFT implementation.

\subsection{PR algorithms}\label{ssec:pralgs}

We evaluated three PR algorithms implemented in the Phase Retrieval Toolbox~(PHASERET, \cite{ltfatnote045}): Phase-gradient heap integration ({PGHI})~\cite{ltfatnote040}, fast Griffin-Lim (FGLA)~\cite{Perraudin2013griffin}, and single-pass spectrogram inversion ({SPSI})~\cite{beauregard2015single}. \reviewed{These algorithms were chosen to cover a wide range of PR strategies and based on the availability of tested and reliable implementations in the toolbox, enabling a fair comparison.} PHASERET relies on LTFAT for STFT computation and other basic \reviewed{functionalities}. 

\textbf{1) PGHI} is a non-interative method. \reviewed{PGHI implies no assumptions on the signal. Instead, }it is based on the phase-magnitude relations of an STFT computed using a Gaussian window~\cite{portnoff1979magnitude}, namely, the relation between the partial phase derivatives of the continuous STFT with a Gaussian window and the partial derivatives of the logarithmic STFT magnitudes~\cite{po76,ltfatnote040}.  \reviewed{In PGHI, this relation is approximated for the discrete STFT as:

  \begin{equation}\label{eq:phase-magnitude}
    \begin{split}
    \partial_n \mathrm{\phi}_g[m,n] & \approx \frac{aM}{\lambda} \partial_m \log(|\operatorname{S}_g|)[m,n],\\
    \partial_m \mathrm{\phi}_g[m,n] & \approx -\frac{\lambda}{aM} \partial_n \log(|\operatorname{S}_g|)[m,n] - 2\pi na/M.
    \end{split}
  \end{equation}
  
Here, $\partial_n,\partial_m$ denote numerical differentiation with respect to $n$ and $m$, respectively. This step may be realized, e.g., by a finite difference scheme. The dependence on properties of the Gaussian STFT suggests a dependence on the window function, which was already observed in~\cite{ltfatnote043}. From the phase-magnitude relation, the phase ($\mathrm{\phi}_g$) is reconstructed in an adaptive integration scheme. The reliance on numerical differentiation and integration suggests that the results of PGHI also depend on the STFT parameters $a,M, \lambda$ and, in particular, its redundancy $D$. In our experiments, PGHI was initialized with no prior knowledge of the original phase.}

\textbf{2) FGLA} is an iterative algorithm relying on alternating projections and it is based on the Griffin-Lim algorithm (GLA)~\cite{griffin1984signal}, which is itself an extension of the seminal Gerchberg-Saxton algorithm~\cite{gerchberg1972practical}. 
Specifically, given a target STFT magnitude combined with an initial phase estimate (\reviewed{in our experiments, the phase was uniformly set to zero}), the algorithm performs first a projection onto the space of consistent STFTs. Since the latter is a strict subspace of $\CC^{M\times N}$ whenever the STFT is redundant, this step is expected \reviewed{to} yield a magnitude different from the target. Therefore, the second step keeps only the new phase, an imposes the target magnitude. Both steps of GLA are repeated until convergence or until a certain number of iterations. A final inverse STFT is then applied to synthesize a  time-domain signal. Thus, GLA does not rely on a signal model, but only on the redundancy of the STFT. \reviewed{In our experiments we employ fast Griffin-Lim (FGLA)~\cite{Perraudin2013griffin}, which adds an acceleration term, governed by a hyperparameter $\alpha$ to the GLA update, and empirically yields better results at the same number of iterations, often significantly. For our experiment we use the default value $\alpha=0.99$ proposed by the reference implementation in LTFAT~\cite{ltfatnote030}. Furthermore, we set the number of iterations to $100$, which provides a decent trade-off between quality and computation time. The convergence curve after $100$ iterations is already rather flat such that a large number of iterations is required to achieve significant improvements.}

\textbf{3) SPSI} is another non-iterative method. In contrast to PGHI, SPSI does not rely on mathematical properties of the STFT. And in contrast to FGLA, it is fast and directly suitable for real-time usage. 
SPSI implicitly assumes a sinusoidal signal model and thus fails for transient and broadband components in the signal. At every time step, SPSI locates peaks in the TF coefficients obtained and predicts the phase by assuming a linear phase progression at the rate of the closest peak frequency. Hence, the rate at which the TF coefficients vary over time is expected to be a limiting factor of PR by SPSI. The phase prediction, being  similar to the integration scheme of PGHI, depends on the time step parameter $a$\reviewed{ and, to a lesser degree, on the number of frequency channels $M$}. \reviewed{In our experiments, SPSI was initialized without any information regarding the original phase.}

\subsection{Evaluation measures}
\label{subsec:evaluation_measures}
\reviewed{
The results were evaluated numerically by means of several measures. First, we considered the signal-to-noise ratio calculated on the magnitude spectrogram, ($\mathbf{SNR_{MS}}$).  \reviewed{This quantity is also sometimes referred to as \emph{spectral convergence}}. Second, to consider human-like performance, we computed the objective difference grade ($\mathbf{ODG}$) based on perceptually motivated models.} 

\textbf{1) Spectrogram signal-to-noise ratio ($\mathbf{SNR_{MS}}$)}
is the logarithmic ratio between the energy of the spectrogram $\abs{S}$ of the original signal $s$ and the energy of the spectrogram difference $(\abs{S_r} - \abs S)$, where $S_r$ is the STFT of the reconstructed time-domain signal $s_r$:

\begin{equation}
\mathbf{SNR_{MS}}(S, S_r) = 10 ~ \log_{10} \frac{\Vert S \Vert ^2}{\Vert  \abs{S_r} - \abs S \Vert ^2}.
\label{eq:SNRformula}
\end{equation}

To compute $\mathbf{SNR_{MS}}$, we used the STFT as in Eq.~\eqref{eq:STFT_final} with $M=2048$, $a=128$ \reviewed{(thus, $D_{SNR}=16$}), and the Gaussian window $g_\lambda$ with $\lambda_\reviewed{{SNR}} = aM/\xi_s \approx 11.886$. In Section \ref{subsec:distorted_phase}, we show that $\mathbf{SNR_{MS}}$ only exhibits minor dependence on this parameter choice. 

\reviewed{
\textbf{2) Objective difference grade ($\mathbf{ODG}$)}
is the overall quality measure introduced in PEAQ~\cite{thiede2000peaq,recommendation20011387} and designed to mimic perceptual quality ratings made by a human listener. PEAQ is a full-reference algorithm, i.e., it performs a direct comparison between a modified signal and a target signal\footnote{In our case, the original signal.}. It relies on an auditory model obtained by processing STFT coefficients and ranges from $0$ to $-4$ with the interpretation shown in Tab.~\ref{tab:ODG}. We used the implementation from~\cite{kabal2002examination}. 

Internally, $\mathbf{ODG}$ computes the STFT of the analysed signal. Thus, it might prefer a particular set of STFT parameters. To consider this effect in our evaluation, we initially calculated ODGs based on PEMO-Q~\cite{huber2006pemo} as well, which uses a Gammatone filterbank. Due to this difference in analysis dictionary, it is unlikely that PEMO-Q exhibits preference for certain STFT parameters. 
We used the implementation from~\cite{emiya2011subjective} and 
refer to this measure as $\mathbf{ODG_{PEMO}}$.
}
\begin{table}[tb]
	\centering
	\begin{tabular}{ll}
		\hline
$\mathbf{ODG}$ & Impairment
\\ \hline
0 & Imperceptible		 \\ 
-1 & Perceptible, but not annoying \\
-2 & Slightly annoying \\
-3 & Annoying\\
-4 & Very annoying \\
 \hline \\
	\end{tabular}
	\caption{Interpretation of $\mathbf{ODG}$.}
	\label{tab:ODG}
	\vspace{-1em}
\end{table}

\section{Experiments}

\subsection{Sensitivity of the evaluation measures}
\label{subsec:distorted_phase}

For 
PR performance, we expected a significant effect of the TF ratio and redundancy across the tested PR algorithms. In order to distinguish between actual effects of the PR algorithm and effects induced by the evaluation measure, we first determined the sensitivity of the evaluation measures $\mathbf{ODG}$, \reviewed{ $\mathbf{ODG_{PEMO}}$,} and $\mathbf{SNR_{MS}}$ to changes of these parameters.


\reviewed{
$\mathbf{SNR_{MS}}$ reflects the average amount of phase distortion over signal duration, thus, we hypothesized that it is largely insensitive to changes in the TF ratio $\lambda$. On the other hand, adjacent TF coefficients are correlated (with the correlation increasing with $D$) and any uncorrelated distortion imposed on the coefficients partially cancels in the synthesis process. Thus, we expected the PR quality to improve with increasing redundancy, and $\mathbf{SNR_{MS}}$ to reflect this. However, $\mathbf{SNR_{MS}}$ itself is based on an STFT, the parametrization of which, determined by $\lambda_{SNR}$ and $D_{SNR}$, might affect the results. To account for this effect, for each evaluated condition, we calculated $\mathbf{SNR_{MS}}$ for all combinations of $\lambda_{SNR} \in \{10^{-3}, 10^4\}$ and $D_{SNR} \in \{2, 8, 32\}$ and evaluated the statistics by means of the average and standard deviation.

The manifestation of phase distortion in synthesized time-domain signals $\tilde{s}$ depends on the width of the synthesis window $\tilde{g}$, which further depends on $\lambda$. Therefore, we expected an effect of the TF ratio on $\mathbf{ODG}$. To account for possible presence effects caused by the STFT parametrization in PEAQ, 
we additionally calculated $\mathbf{ODG_{PEMO}}$ for the same set of conditions.
}

\begin{figure*}[t!]
    \centering
    \includegraphics[width=0.31\linewidth]{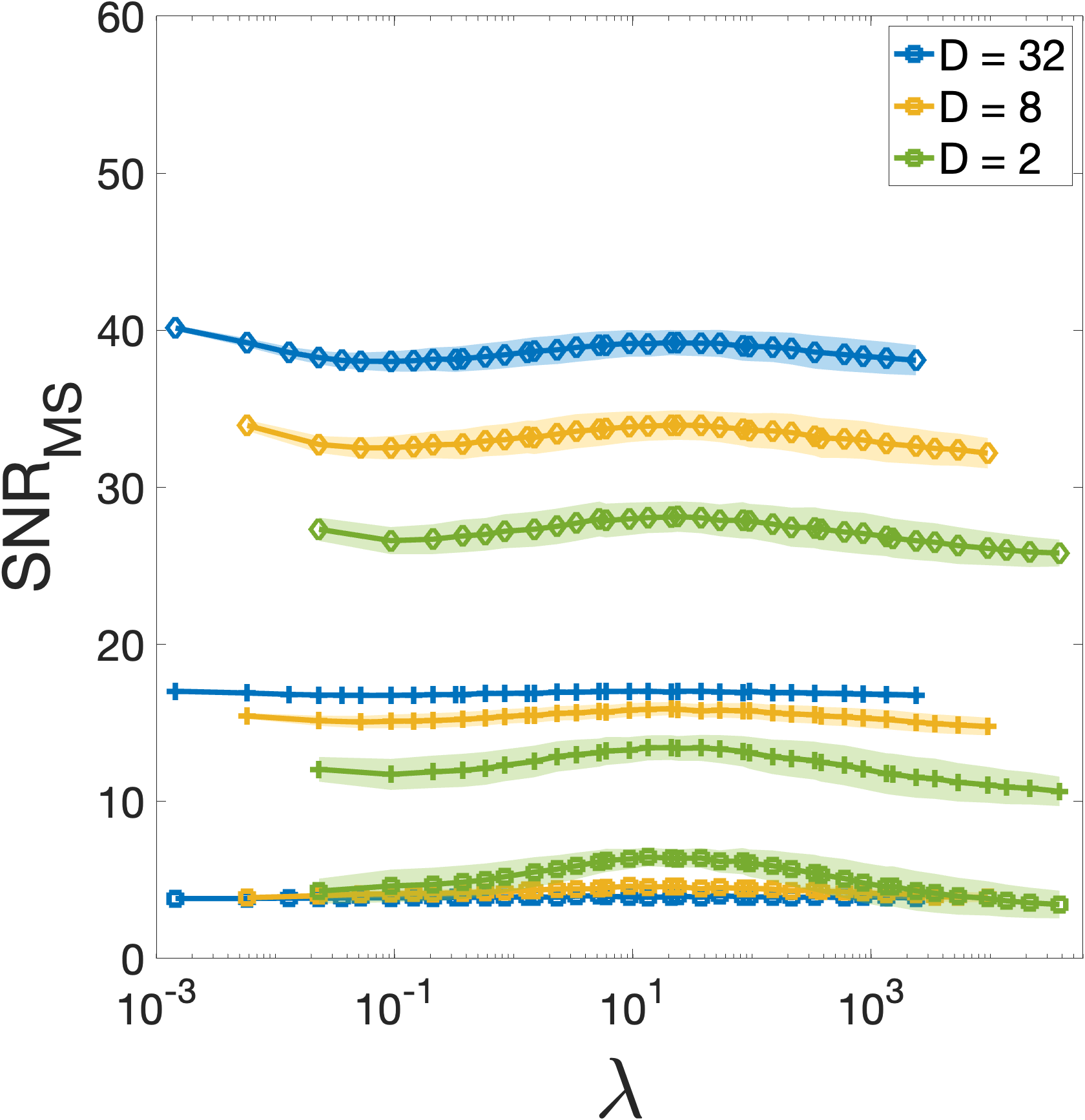}
    \includegraphics[width=0.303\linewidth]{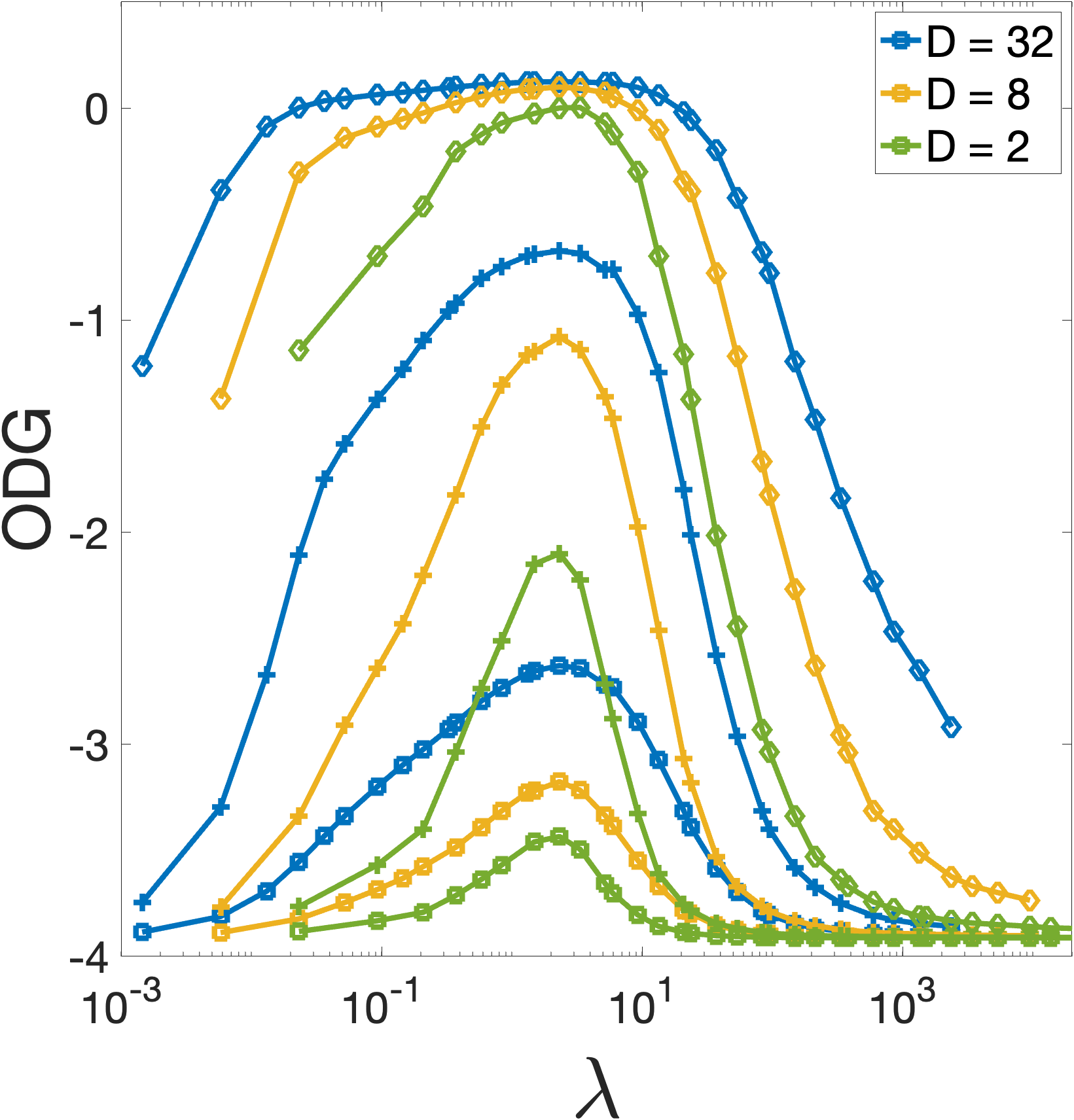}
    \includegraphics[width=0.31\linewidth]{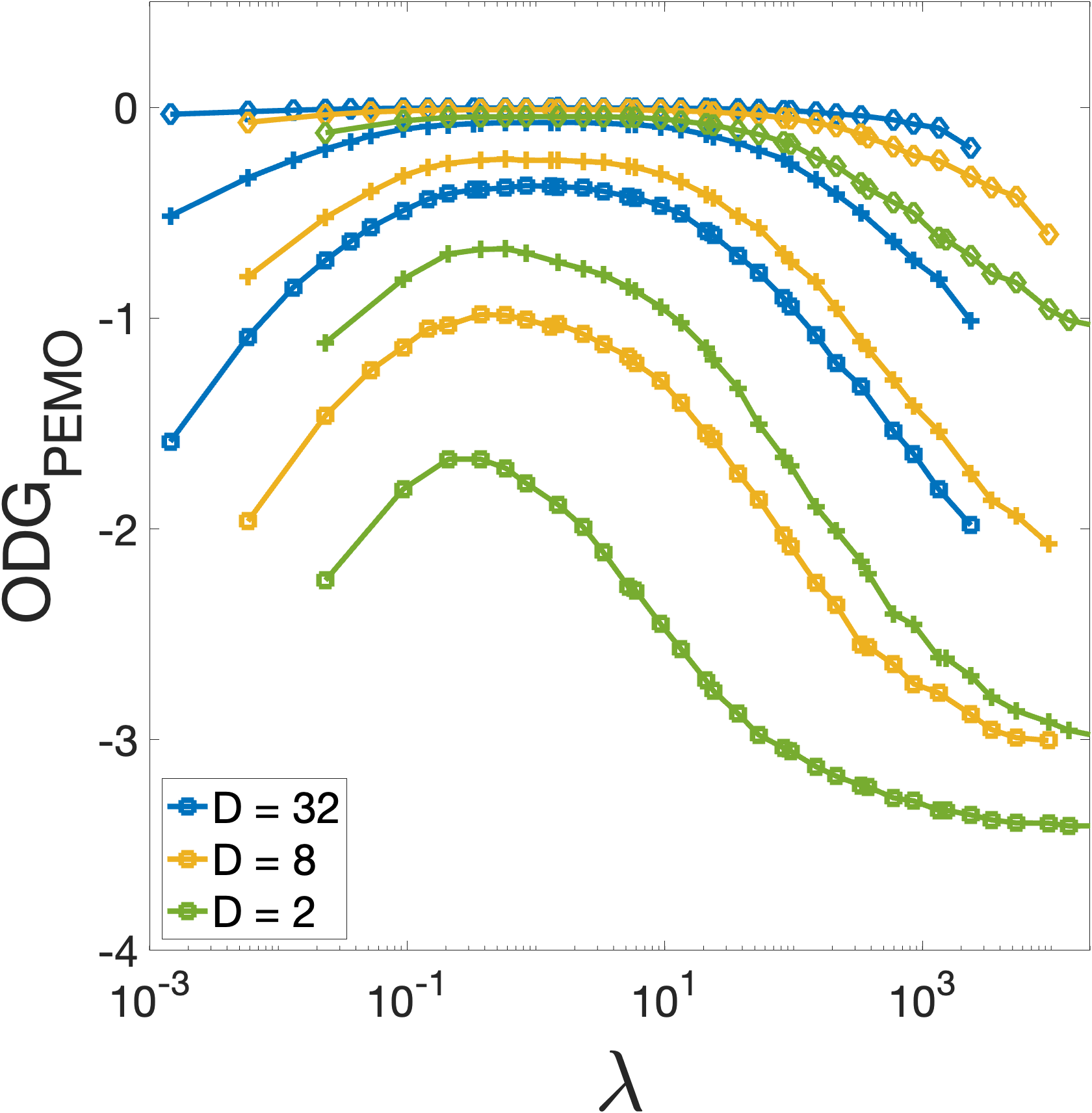}
    \caption{\reviewed{$\mathbf{SNR_{MS}}$, $\mathbf{ODG}$, and $\mathbf{ODG_{PEMO}}$ resulting from the inverse STFT of speech signals with distorted phases. Symbols: severity of distortion with $\sigma=0.1$ (diamonds), $\sigma=0.5$ (plus), $\sigma=1.0$ (squares).}}
    \label{fig:distorted}
\end{figure*}

For the evaluation, we first computed STFTs for various TF ratios $\lambda \in \{10^{-3}, 10^4\}$ and three redundancies $D \in \{2, 8, 32\}$. \reviewed{Note that a $\lambda, D$ combination uniquely determines both $a$ and $M$.} Then, we added Gaussian white noise to the phase of these STFTs. We tested three standard deviations $\sigma \in \{1$, $0.5$, $0.1\}$ and phase values were wrapped onto the range $\pm \pi$ after applying the distortion. \reviewed{Further note that the condition with the highest distortion level corresponds to reconstructing spectrograms with a nearly random phase}. Finally, we calculated the inverse STFT and applied the three measures to the result. This setup provides direct evidence for the extent of sensitivity of $\mathbf{ODG}$, $\mathbf{ODG_{PEMO}}$, and $\mathbf{SNR_{MS}}$ to the TF ratio $\lambda$ and redundancy $D$.

Fig. \ref{fig:distorted} shows the results. \reviewed{For the $\mathbf{SNR_{MS}}$, the statistics across $\lambda_{SNR}$ and $D_{SNR}$ show negligible standard deviation as compared to the effect of changes of $\lambda$ and $D$. This indicates that the parametrization of 
$\mathbf{SNR_{MS}}$ had nearly no effect on the sensitivity and reliability of this measure when evaluating phase effects for various $\lambda$ and $D$. The average $\mathbf{SNR_{MS}}$ correspond roughly to the values calculated for $D_{SNR}=16$ and $\lambda_{SNR}=10$. Thus, in the following, we use parameters as indicated in Sec.~\ref{subsec:evaluation_measures} to calculate $\mathbf{SNR_{MS}}$ in the following experiments.} 

For the highest distortion level ($\sigma = 1$), the $\mathbf{SNR_{MS}}$ was below 8 dB and \reviewed{$\mathbf{ODG}$ was} worse than `annoying' in most cases, indicating that this amount of distortion substantially destroyed the original signals in every considered combination of $\lambda$ and $D$. These results reflect the output one would expect of our measures when reconstructing spectrograms with a random phase. \reviewed{On the other side, $\mathbf{ODG_{PEMO}}$ was between `not annoying' and `annoying' for the majority of conditions, implying that even for nearly random phases, the results would have been acceptable when relying on this measure alone. This indicates that $\mathbf{ODG_{PEMO}}$ is most probably not suitable as a measure to analyze the effects of phase distortion.}

For the moderate ($\sigma = 0.5$) and low ($\sigma=0.1$) distortion levels, a pattern emerges: at fixed TF width ratio, larger $D$ yielded better performance in terms of larger $\mathbf{SNR_{MS}}$ and better $\mathbf{ODG}$. \reviewed{With most results being better than `not annoying', $\mathbf{ODG_{PEMO}}$ showed little sensitivity to these amounts of noise, except at extreme $\lambda$ and low redundancy $D$. We take this as further evidence to the poor sensitivity of this measure to phase distortion.
}
$\mathbf{SNR_{MS}}$ showed little effect of TF ratio, with a small peak at $\lambda$ of approximately 10. 
In contrast, $\mathbf{ODG}$ seems to be more sensitive to the TF ratio, following a bell shape with a clear peak at the same single-digit $\lambda$ for all $D$s. This peak seems to be wide for low levels of phase distortions and to become sharper for increasing distortion level.

In summary, we observed the following: \reviewed{1) the small differences, when computing $\mathbf{SNR_{MS}}$ parametrized to various $\lambda_{SNR}$ and $D_{SNR}$ combinations, indicate that the particular choice of its own parametrization is not essential.} 2) $\mathbf{SNR_{MS}}$, \reviewed{calculated with the default parametrization,} is sensitive to phase distortions, but for a fixed amount of distortion, it is only mildly sensitive to the TF ratio. Moreover, at low to moderate distortion, $\mathbf{SNR_{MS}}$ increases with $D$.
$\mathbf{SNR_{MS}}$ below 10 dB was in line with what we expect for reconstructions with a random phase and can be used as a rule of thumb when evaluating PR algorithms. 2) $\mathbf{ODG}$ is sensitive to both TF ratio and redundancy, however, it showed ceiling effects, i.e., it saturated at low levels of distortions and high redundancies. The sensitivity to the TF ratio seems to depend on the level of distortion, with single-digit $\lambda$ being a good choice at most redundancies. \reviewed{ 3) $\mathbf{ODG_{PEMO}}$ was barely sensitive to phase distortions and did not use the full range even for nearly random phases. This provides strong indication that it is not a suitable measure for evaluating phase effects.} 4) All measures showed consistently better results with increasing redundancy, demonstrating the increased robustness of the inverse STFT to phase distortions with increasing redundancy. 5) By combining $\mathbf{SNR_{MS}}$ and $\mathbf{ODG}$ we can avoid our evaluation to be hampered by 
the low sensitivity of $\mathbf{SNR_{MS}}$ to $\lambda$ and the ceiling effects of $\mathbf{ODG}$ for high $D$, obtaining an adequate scheme for evaluating PR algorithms in the following experiments. 

\subsection{Effect of STFT parameters on PR}
\label{subsec:comp_retrieval_algorithms}

In this experiment, we studied the effect of the choice of STFT parameters on PR in terms of $\mathbf{SNR_{MS}}$ and $\mathbf{ODG}$. Evaluation was performed for Gaussian windows, while varying the redundancy $D$, and TF ratio $\lambda$. The experiment aims to not only assess the effect of $D$ and $\lambda$, but also to demonstrate performance differences between the algorithms. 

\begin{figure}
    \centering
    \includegraphics[width=0.98\columnwidth]{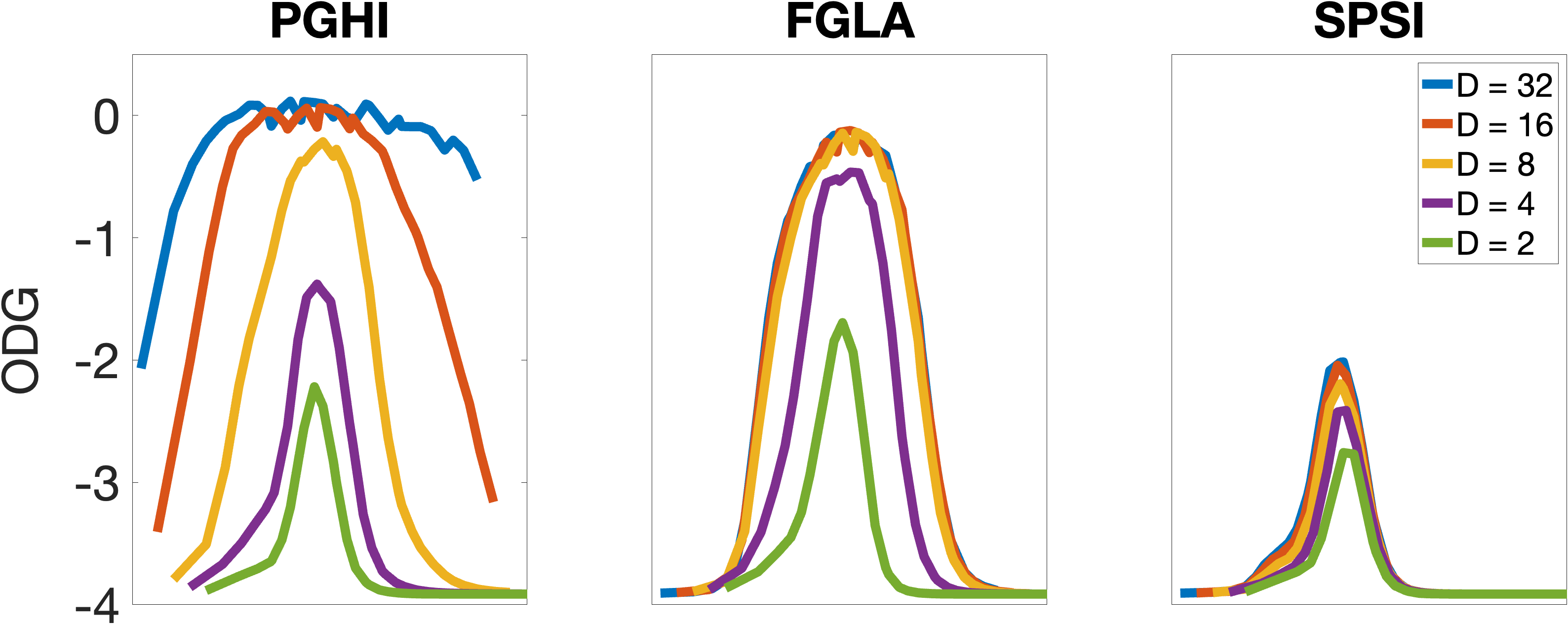} 
    \includegraphics[width=0.98\columnwidth]{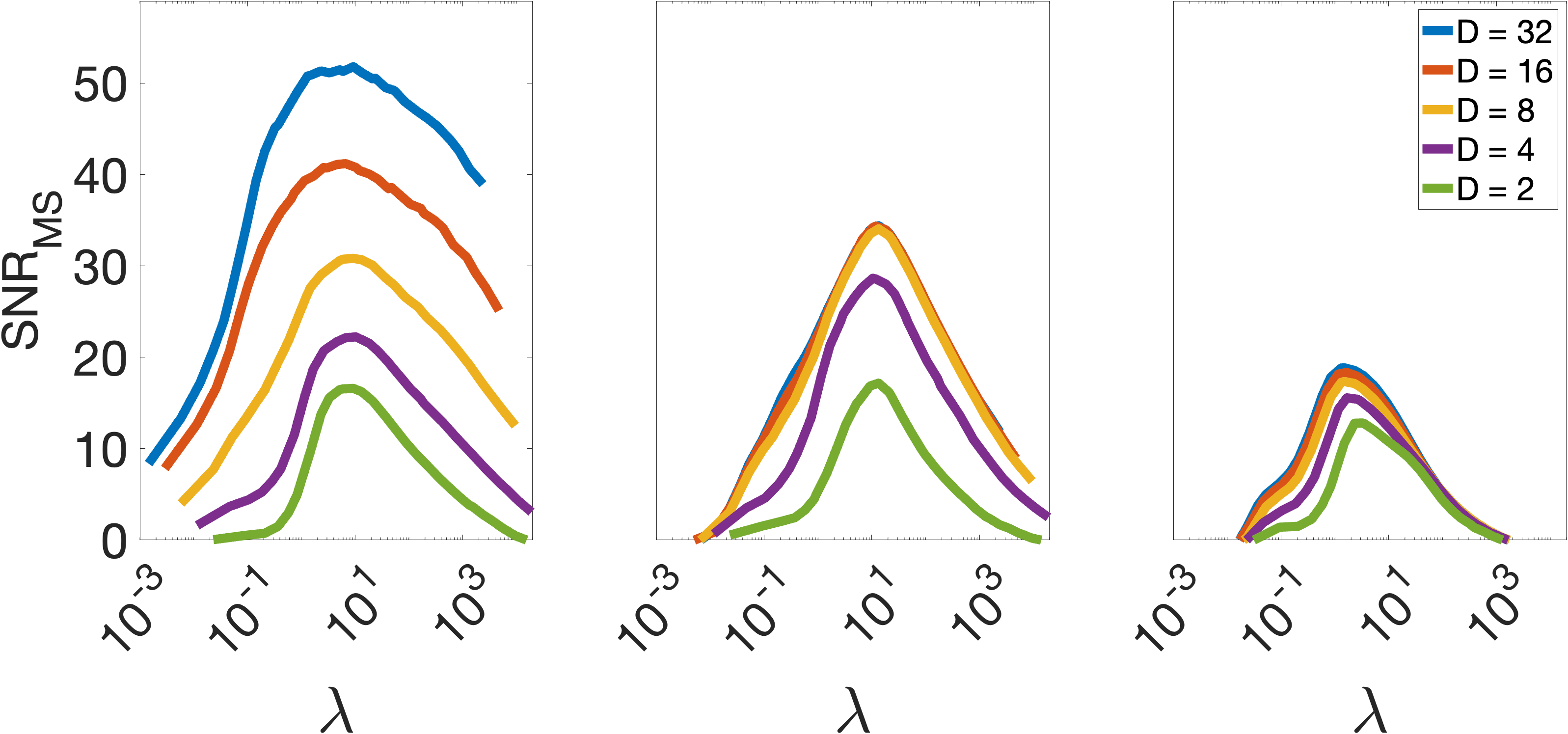}
    \caption{\reviewed{PR performance in terms of $\mathbf{ODG}$ and $\mathbf{SNR_{MS}}$ obtained with three PR algorithms: PGHI, SPSI, and FGLA. Calculations done with the Gaussian window, and various redundancies and TF ratios.}}
    \label{fig:full_Gauss}
\end{figure}


To achieve this, we created spectrograms of the speech dataset considering redundancies $D\in\{2,4,8,16,32\}$ and a large range of TF ratios $\lambda \in [10^{-3}, 10^4]$, applied all considered PR algorithms (Sec. \ref{ssec:pralgs}) on those spectrograms, and calculated $\mathbf{SNR_{MS}}$ and $\mathbf{ODG}$ of the reconstructed signals. The results are presented in Fig.~\ref{fig:full_Gauss}. 

For all three algorithms, $\lambda$ had a clear effect on both $\mathbf{SNR_{MS}}$ and $\mathbf{ODG}$. We can clearly see that the change of $\mathbf{ODG}$ in $\lambda$ is more pronounced than in Exp. A, indicating actual impact of $\lambda$ on PR quality. In contrast to Exp. A, $\mathbf{SNR_{MS}}$ also shows significant variation in $\lambda$, providing further evidence that this is the case. For $\lambda$ below 0.1 and above 100, the level of phase distortions introduced by the PR corresponded to that of high level of noise ($\sigma = 1.0$). Although the general trend is similar across all algorithms, the extent to which STFT parameters affect PR quality depended on the chosen algorithm.

For PGHI, both measures showed peaks at $\lambda$, and those peaks were shared for every $D$. For PGHI peaks were less pronounced than for FGLA, indicating that PGHI is less sensitive to a particular choice of $\lambda$. For $\mathbf{ODG}$, the peak was at $\lambda = 2.32$, corresponding to $M=320$ for $D=2$. For $\mathbf{SNR_{MS}}$, the peak was at $\lambda = 5.94$, corresponding to $M=512$ for $D=2$. The performance increased with the redundancy, showing ceiling effects in $\mathbf{ODG}$ for redundancy of 16 or larger. For redundancies of $D\geq 16$, $\mathbf{SNR_{MS}}$ showed little distortions, comparable to our low distortion level with Gaussian white noise. From these results, we conclude that for speech signals PGHI works best at $D \geq 16$ with $\lambda \approx 3$. For lower redundancies, the performance degraded and the choice of $\lambda$ became even more important.

For FGLA, both measures followed a thin bell shape with peaks at $\lambda$. This peak was the same at every redundancy $D$. For $\mathbf{ODG}$, the peak was at $\lambda = 3.34$, corresponding to $M=384$ for $D=2$. For $\mathbf{SNR_{MS}}$, the peak was at $\lambda = 13.37$, corresponding to $M=768$ for $D=2$. For both measures, the performance increased with the redundancy for D of up to 8, showing no improvements beyond that redundancy. This is in contrast with Sec. \ref{subsec:distorted_phase}, where performance improved with increasing redundancy. We conclude that for FGLA, the choice of $\lambda$ is crucial, with $\lambda \approx 8$ providing good results. For redundancies $D \leq 4$, FGLA performed better than PGHI, but, there is no gain in increasing $D$ beyond $8$. 

For SPSI, even the best performance corresponded to large or moderate level of distortion when compared to that obtained for Gaussian white noise from Sec. \ref{subsec:distorted_phase}. Performance increased slightly with the redundancy, however still remained at a low level. The performance depended on $\lambda$, showing a peak $\lambda=1.4$ for every $D$, corresponding to $M=256$ at $D=2$. The low general performance of SPSI is probably an effect of the underlying signal assumption of slowly varying sinusoidal components, which does not hold for speech signals. 

In all three algorithms, the performance increased with the redundancy. This is not surprising because with increasing redundancy, phase and magnitude are more dependent and deducing one from the other becomes easier. This is related to the fact that phase can be perfectly calculated from the magnitude of a continuous STFT (up a fixed scalar factor and disregarding numerical precision)~ \cite{portnoff1979magnitude} and, by increasing redundancy, the discrete STFT approximates the continuous setting. Algorithms utilizing this principle particularly benefit from increased redundancy. This explains the performance increase with redundancy provided by PGHI (which is based on that principle) and the limited performance gain provided by FGLA and SPSI (which rely on other principles). 

The time-frequency ratio $\lambda$ affected the performance of all tested algorithms. While the redundancy extends the range of reasonable choices for $\lambda$, generally good performance can be obtained for $\lambda$ between 0.2 and 20, whereas performance was mostly poor for $\lambda$ below 0.1 and above 100. In order to explain why particular time-frequency ratios are beneficial for PR, we need to look into the interaction between the audio signal and the window duration resulting from a particular $\lambda$. For example, at this sampling rate, $\lambda=5$ implies a window duration of approximately 35~ms and provides a good trade-off between temporal and spectral resolution, see Fig. \ref{fig:spec_samered}. Substantially shorter and longer windows, i.e., resulting from substantially smaller and larger time-frequency ratios, respectively, create more spectral and temporal smearing in the magnitudes, respectively. 

\subsection{Effect of STFT parameters on inconsistent spectrograms}

In the previous experiment we considered PR from unmodified magnitude spectrograms. This allowed us to investigate the PR task in isolation, without having to consider inconsistency, e.g., introduced in processing. However, in practical applications, PR is mostly applied to modified or synthetic spectrograms, which are rarely consistent.
Reconstruction from inconsistent spectrograms introduces errors, recall Fig. \ref{fig:consistency}. When combined with PR, these errors cannot be uniquely attributed to either inconsistency or PR artifacts. Hence, we investigated the PR effect on inconsistent spectrograms in a simple setting: Approximating a  time-invariant filter with nonnegative frequency response by weighting spectrogram channels according to the (sampled) frequency response and subsequent PR of the phaseless weighted spectrogram. The target signal is obtained by applying the frequency response directly to the DFT of the full input signal. As a reference, we use reconstruction from the weighted complex STFT, i.e., we use the original input phase. Given the poor performance of SPSI in Exp. B, we did not consider it for this experiment. The frequency response of the considered filter is

\begin{equation}
    \max(\{\min (\{ 0.1 + \cos(2\pi\cdot 5 \xi), 1\}), 0.1\}), 
\end{equation}
with frequency index $\xi\in\{0,\ldots,L-1\}$.
The filter was chosen only for the purpose of illustration. The filter had the same length as the signal and possessed 15 equidistant peaks where the output's energy was unmodified, and 14 valleys where it was reduced by a factor of 10. The sampled filter resolved all valleys and peaks from the original filter for $M\geq 96$. In this setting, the larger the $M$, the milder the inconsistencies in the processed STFTs. 

\begin{figure}
    \centering
    \includegraphics[width=0.99\columnwidth]{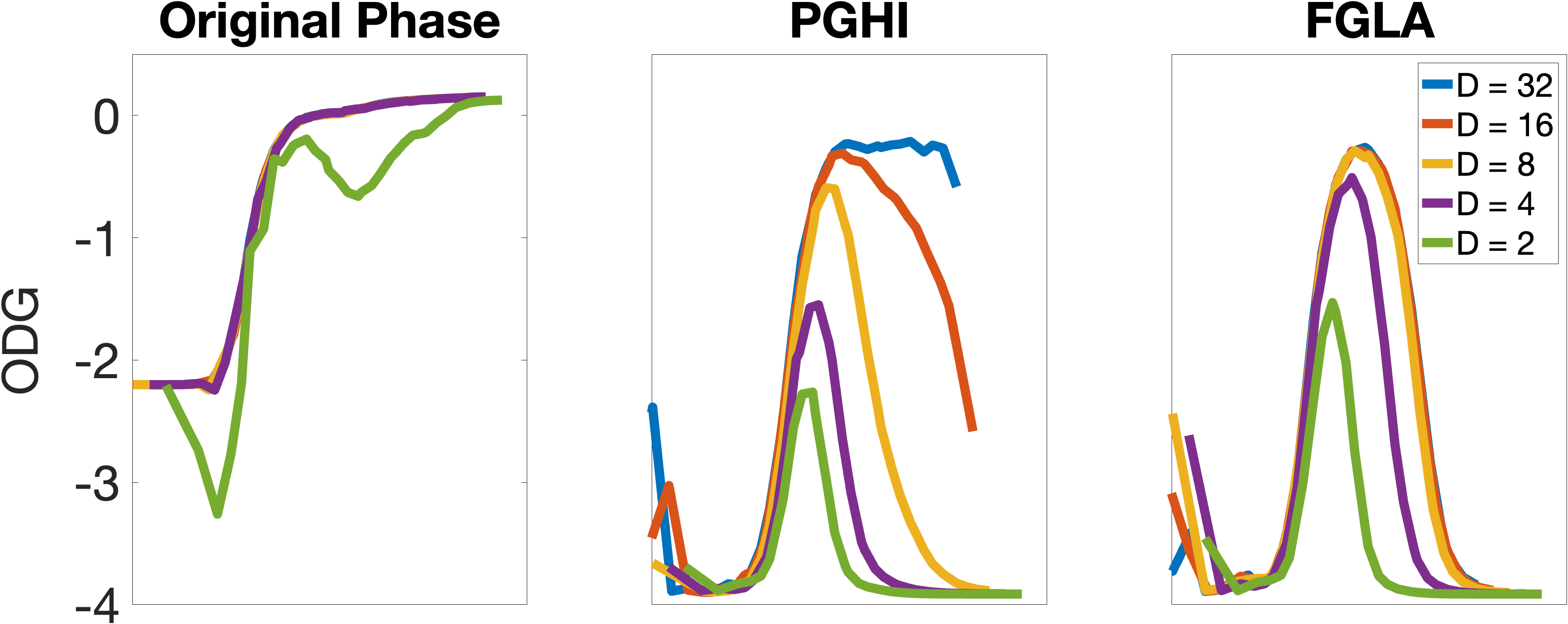}
    \includegraphics[width=0.99\columnwidth]{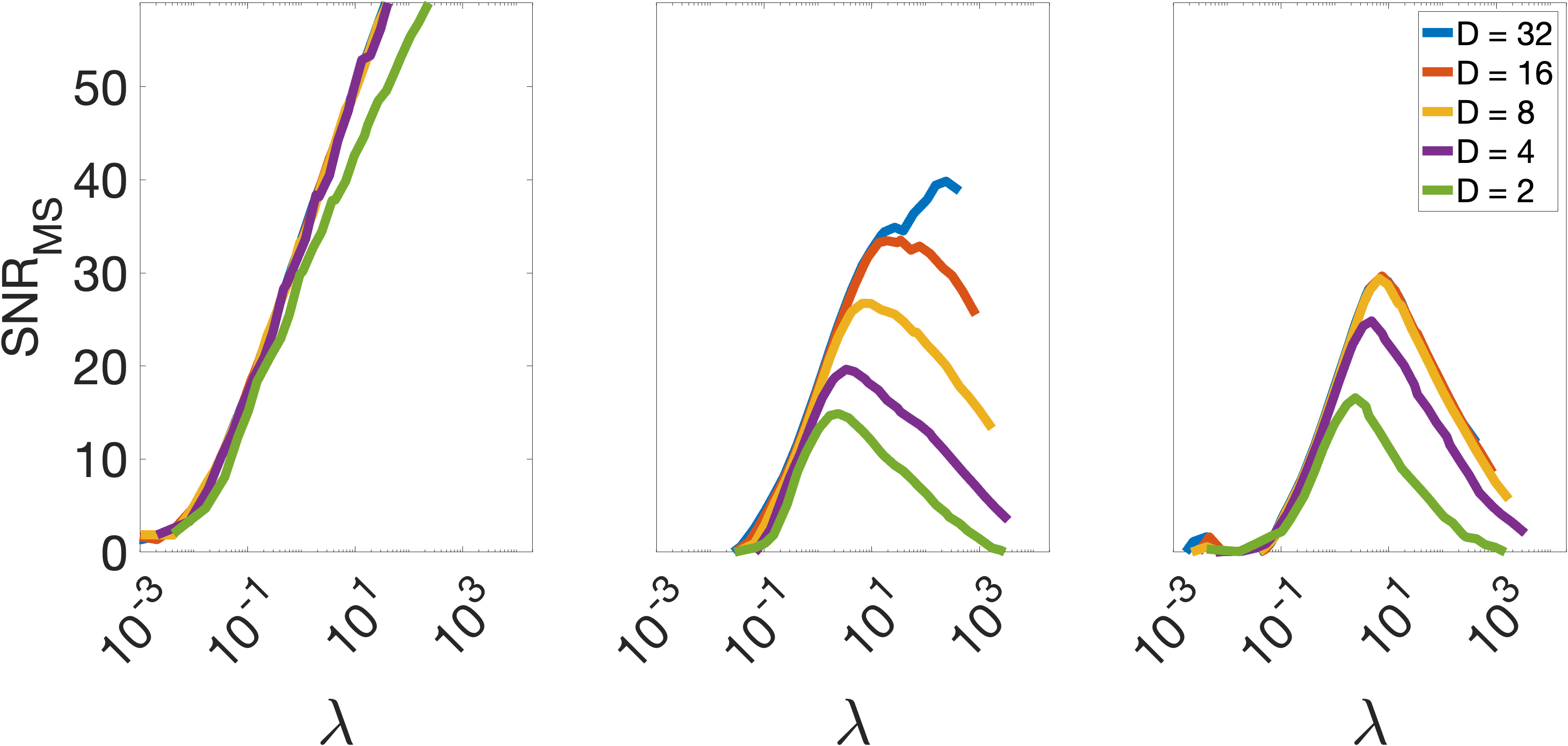}
    \caption{\reviewed{Effect of the synthesis of inconsistent spectrograms in terms of $\mathbf{ODG}$ and $\mathbf{SNR_{MS}}$ obtained with original phases (left columns) and reconstructed phases by PGHI (center columns) and FGLA (right column). Calculations done with the Gaussian window and various redundancies and TF ratios.}}
    \label{fig:inconsistent_magnitudes}
\end{figure}

The results are shown in Fig. \ref{fig:inconsistent_magnitudes}. Using the original phase, performance increased monotonically with $\lambda$. This corroborates our initial hypothesis that the larger the $M$, the less inconsistencies are introduced. The performance obtained by the three PR algorithms also increased with $\lambda$ until a tipping point where their performance decreased similarly to the decrease found in the previous experiment. 

We conclude that the inconsistencies in spectrograms interact with the effect of STFT parameters on PR. This is clear when comparing to Exp. B, where the optimal performance was found for a similar  $\lambda$ at every redundancy. In contrast, in this experiment, where the phase is reconstructed from inconsistent spectrograms, the optimal $\lambda$ increased by a factor of ten when doubling the redundancy.

\subsection{Effect of the signal content}
\label{subsec:comp_signal_content}

In this experiment, we investigate the relationship between PR performance and the signal content. Based on the explanation for the optimal PR parameters from~\ref{subsec:comp_retrieval_algorithms}, we expected the signal content  to affect the optimal PR parameters. To examine this, we performed a reduced version of Exp. B on three classes of audio signals and analyzed their effect on the PR performance. Given that $\lambda$ determines the TF resolution trade-off and uncertainty, we expected to find optimal $\lambda$ depending on the signal content. 

\begin{figure}
    \centering
    \includegraphics[width=0.98\columnwidth]{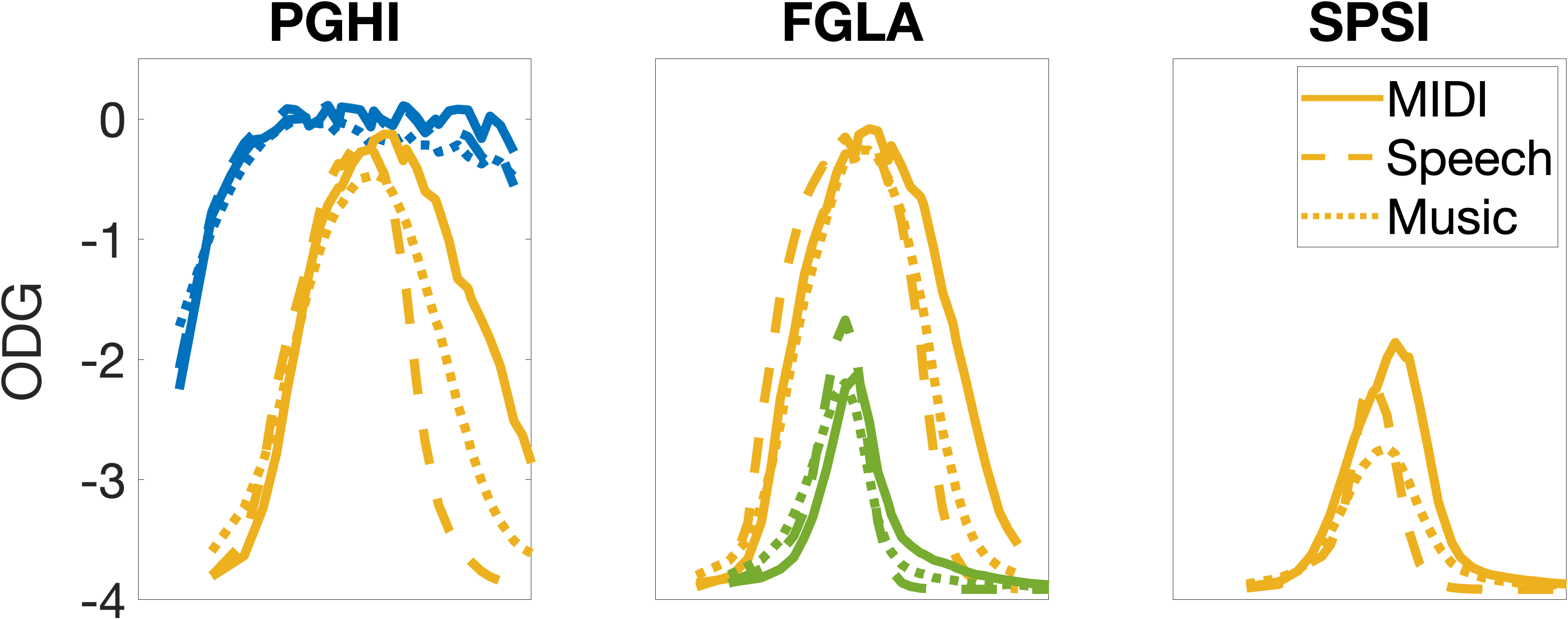}
\\
    \vspace{4pt}
    \includegraphics[width=0.98\columnwidth]{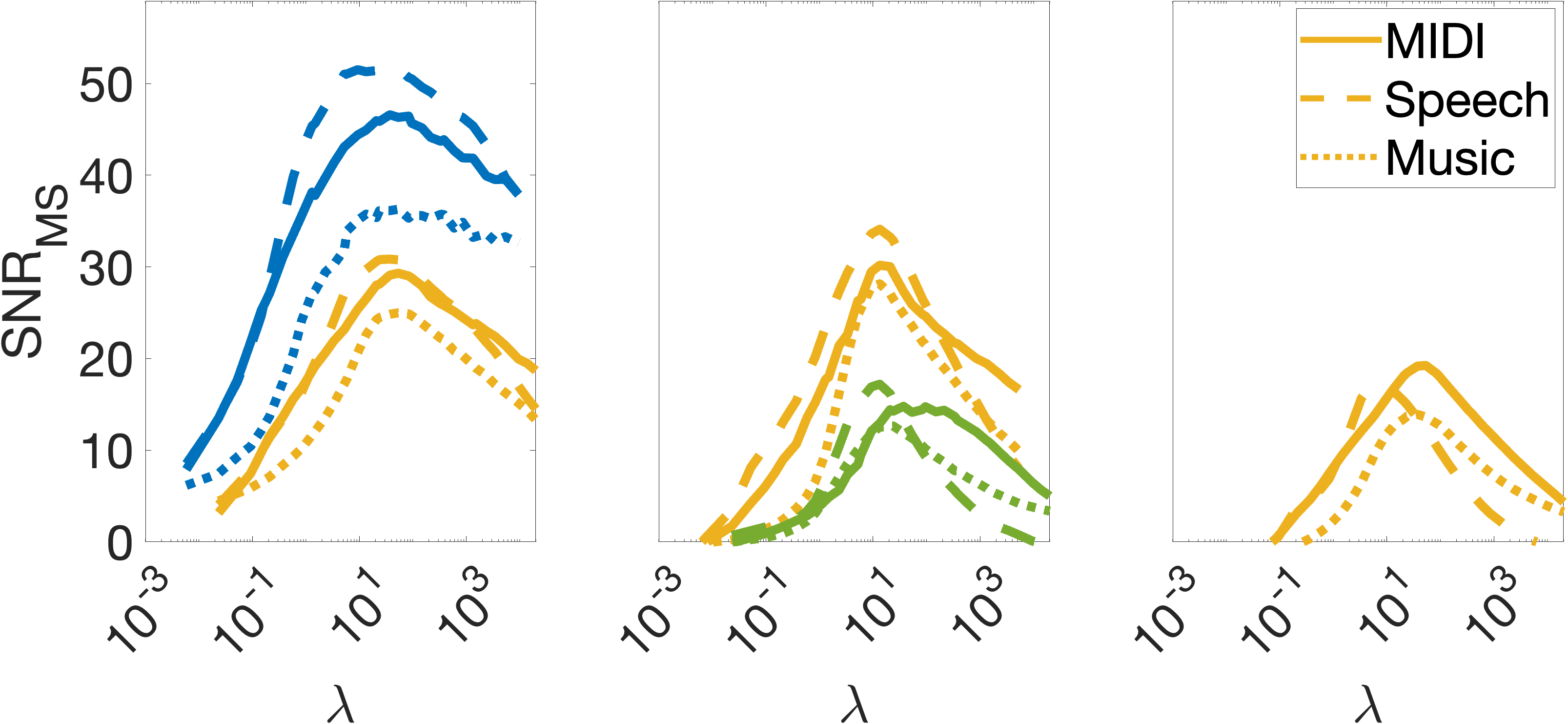}
    \caption{PR performance as an effect of the signal class. MIDI represented in solid lines, speech in dashed lines, and music in dotted lines. Color indicates redundancy: Blue ($D=32$), yellow ($D=8$), and green ($D=2$). All other aspects \reviewed{are} as in Fig.~\ref{fig:full_Gauss}.}
    \label{fig:data_dependencies}
\end{figure}

\reviewed{PR} was performed by all three algorithms \reviewed{on} each of the three considered datasets. For PGHI, we used three redundancies $D\in\{2, 8, 32\}$, for FGLA two redundancies $D\in\{2, 8\}$, and for SPSI only $D=8$, considering the reduced significance of redundancy for FGLA and SPSI in previous experiments. Figure \ref{fig:data_dependencies} shows the results.

The PR performance depended on the signal class, however, it not as strong as the dependency on the STFT parameters. For example, for PGHI and redundancy of $D=32$, we found an average difference in $\mathbf{SNR_{MS}}$ of approximately 10 dB between speech and music, with a largest difference of 20 dB for $\lambda=10$. Such a difference is smaller than that when switching from $D=32$ to $D=8$ or varying $\lambda$ by one order of magnitude. The other PR algorithms showed even smaller effect of the signal class. 

Still, the signal class had a general effect on the optimal $\lambda$, i.e., compared to speech, the optimal $\lambda$ increased by factor three on average when switching to MIDI, with the optimal $\lambda$ for music being between those found for speech and MIDI. This effect can be (again) explained by looking at the interaction between the audio signal and the window duration linked with the time-frequency ratio. Music contains generally lower frequencies, i.e., down to 20~Hz, while speech has the lowest fundamental frequency around 80~Hz. Compared to the MIDI-generated piano sounds, our electronic music had a larger bandwidth. The MIDI-generated piano sounds, on the other hand, contained on average more energy in the low frequency region. Thus, our MIDI required longer window durations, i.e., larger time-frequency ratios, in order to encode the phase information in the magnitude at the same level of accuracy as for speech. 

The general conclusion from this experiment is that 1) PR performance depends only a little on the signal class as compared to the dependency on the STFT parameters, and 2) the optimal $\lambda$ depends on the frequency content to be processed, with lower for speech and higher by up to a factor of three when applied to low frequency signals such as music.

\subsection{Effect of the window function}\label{subsec:windows}


\begin{figure}
    \centering
    \includegraphics[width=0.98\columnwidth]{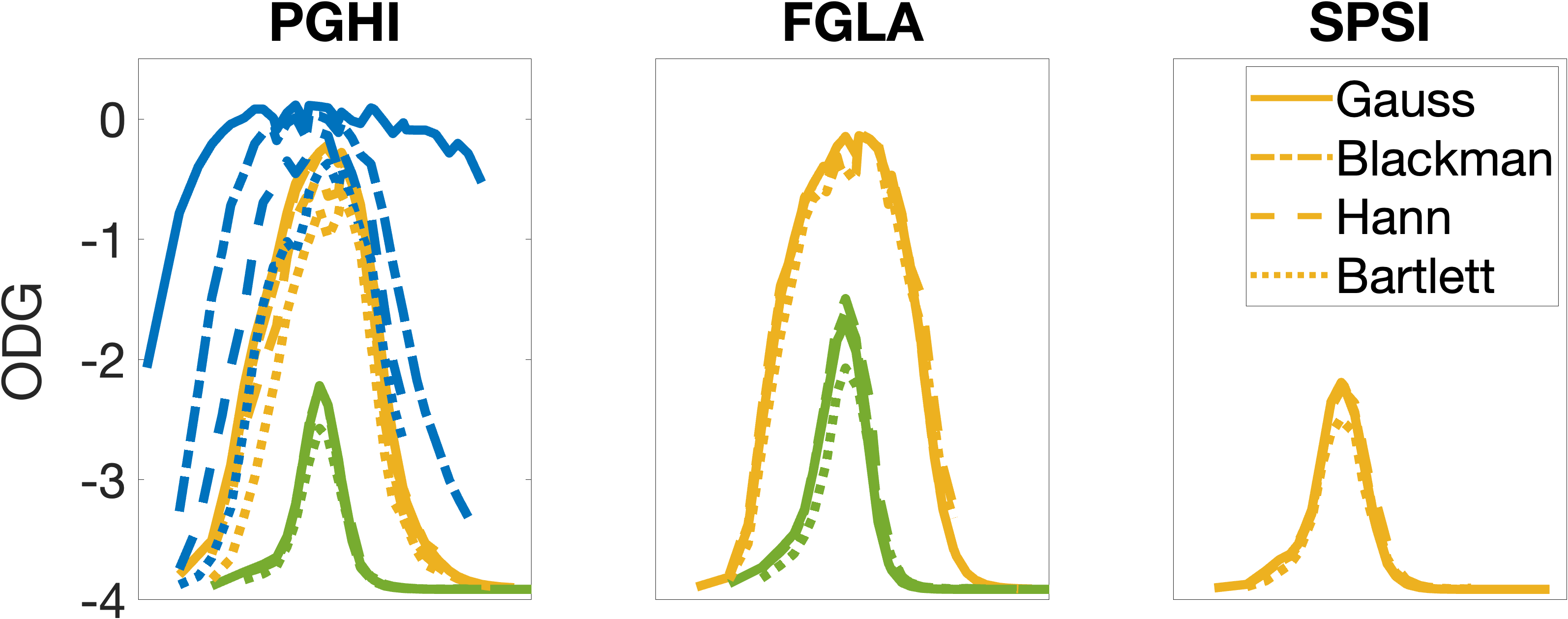}\\
    \vspace{4pt}
    \includegraphics[width=0.98\columnwidth]{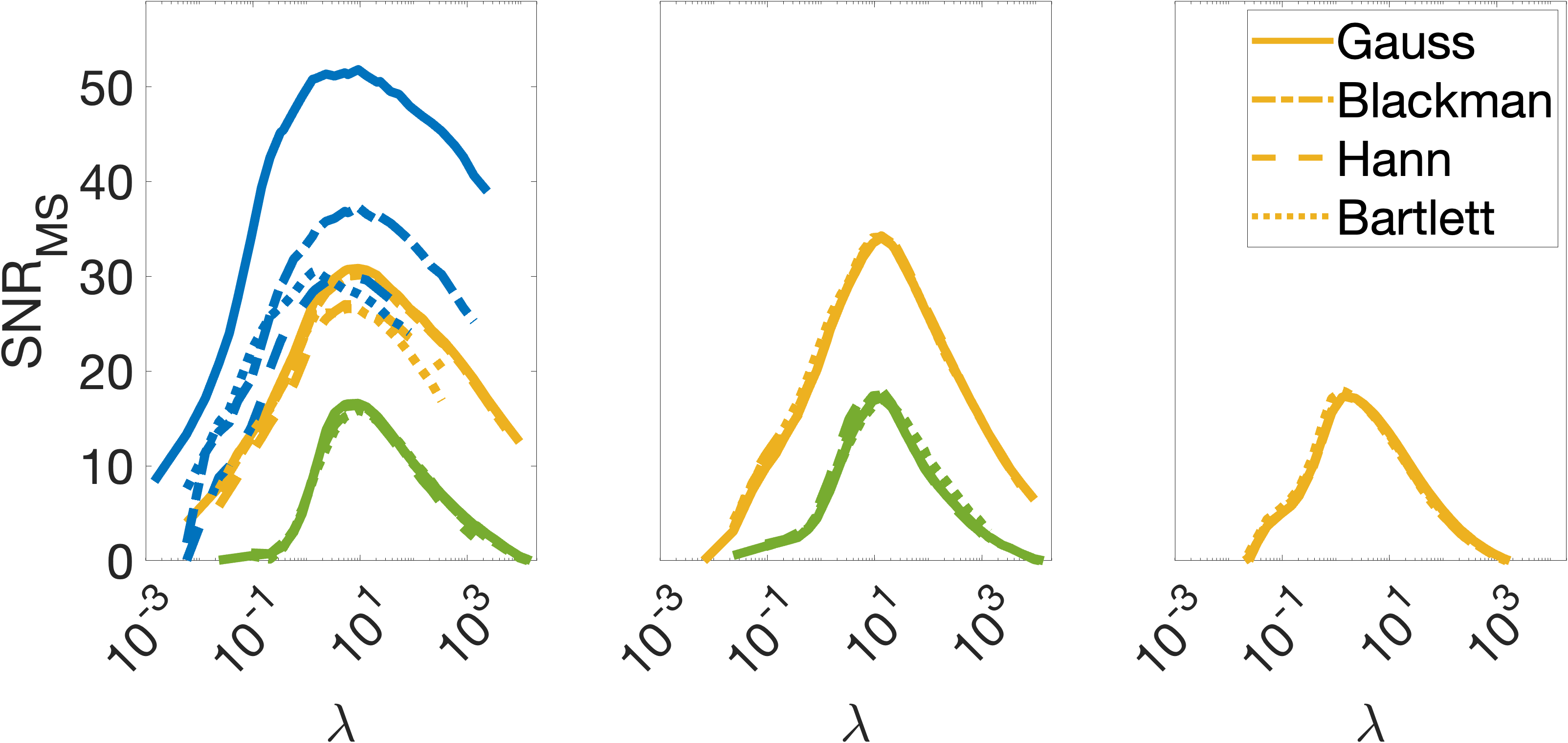}\vspace{1pt}
    
    \caption{\reviewed{PR performance as an effect of the window. Gauss window represented in solid lines, Blackman in dashed lines, Hann in dotted lines, and Bartlett in dash-dot line.  All other aspects are as in Fig.~\ref{fig:data_dependencies}}}
    \label{fig:windows}
\end{figure}

From the three PR algorithms, only PGHI places explicit assumptions on the STFT parameters, specifically the window being a Gaussian.  Therefore, we expect a particular influence of the window function for PGHI. FGLA and SPSI, on the other hand, make little assumptions on the transform, so we expect no large effect of the window function.

To verify these hypotheses, this experiment repeats the Exp. B, 
with the difference that we used \reviewed{either the Gaussian, the Blackman, the Hann, or the Bartlett window in the STFT computations. To match the windows to $\lambda$, we determine $g$ closest to the Gaussian $g_\lambda$, as discussed in Sec. \ref{subsec:resolution_stft}. Following this, we completed the procedure from Exp. B for a comparable range of TF ratios. We evaluated this experiment for all redundancies, but only show results for the same redundancies as in Exp. D in order for the results to be easier to interpret. The results are presented in Fig. \ref{fig:windows}.
}

\reviewed{Only PGHI showed a significant sensitivity to the window. In particular, at a high redundancy such as 32, the difference between every window is significant, even larger than the difference for signal content. At this redundancy, the Gaussian window clearly outperforms every other window, with the Bartlett window performing even worse than the Gaussian window at redundancy 8. For both PGHI and FGLA, the effect of the window was not significant and was well below the effect for signal content. From this experiment we conclude that the choice of window does not significantly affect PR algorithms which do not rely on particular structures of the STFT.}

\subsection{Effect of the convergence of FGLA}

\label{subsec:comp_convergence}
A major drawback of iterative PR algorithms is the necessity to perform multiple time-consuming iterations. In the previous experiments, we were looking for the optimal TF ratio $\lambda$ and used 100 iterations of FGLA in all comparisons. However, there might be an interaction between the TF ratio $\lambda$ and the performance per iteration, yielding a different optimum range $\lambda$ at different number of iterations.

\begin{figure}
    \centering
    \includegraphics[width=0.33\columnwidth]{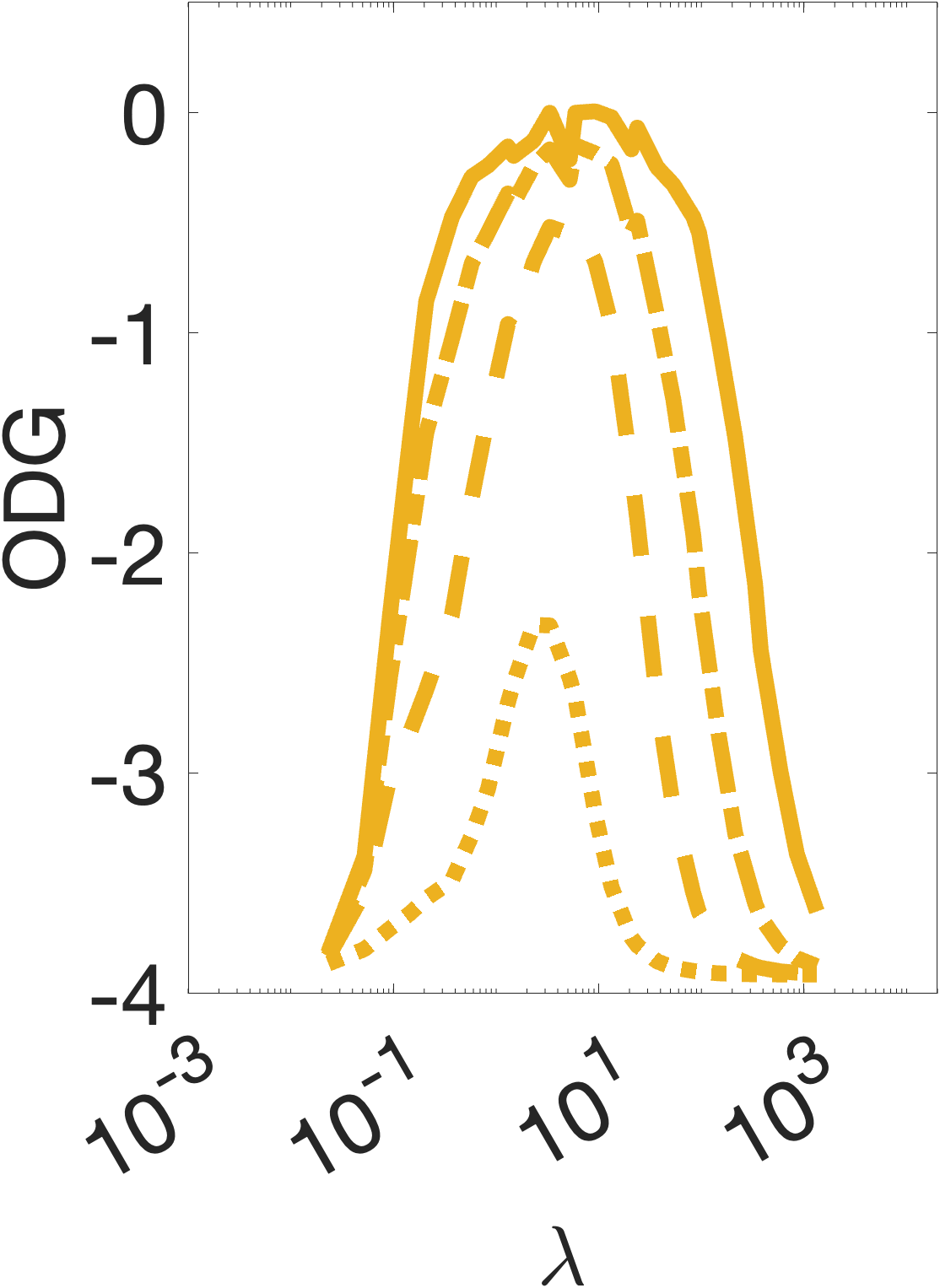}
    \includegraphics[width=0.33\columnwidth]{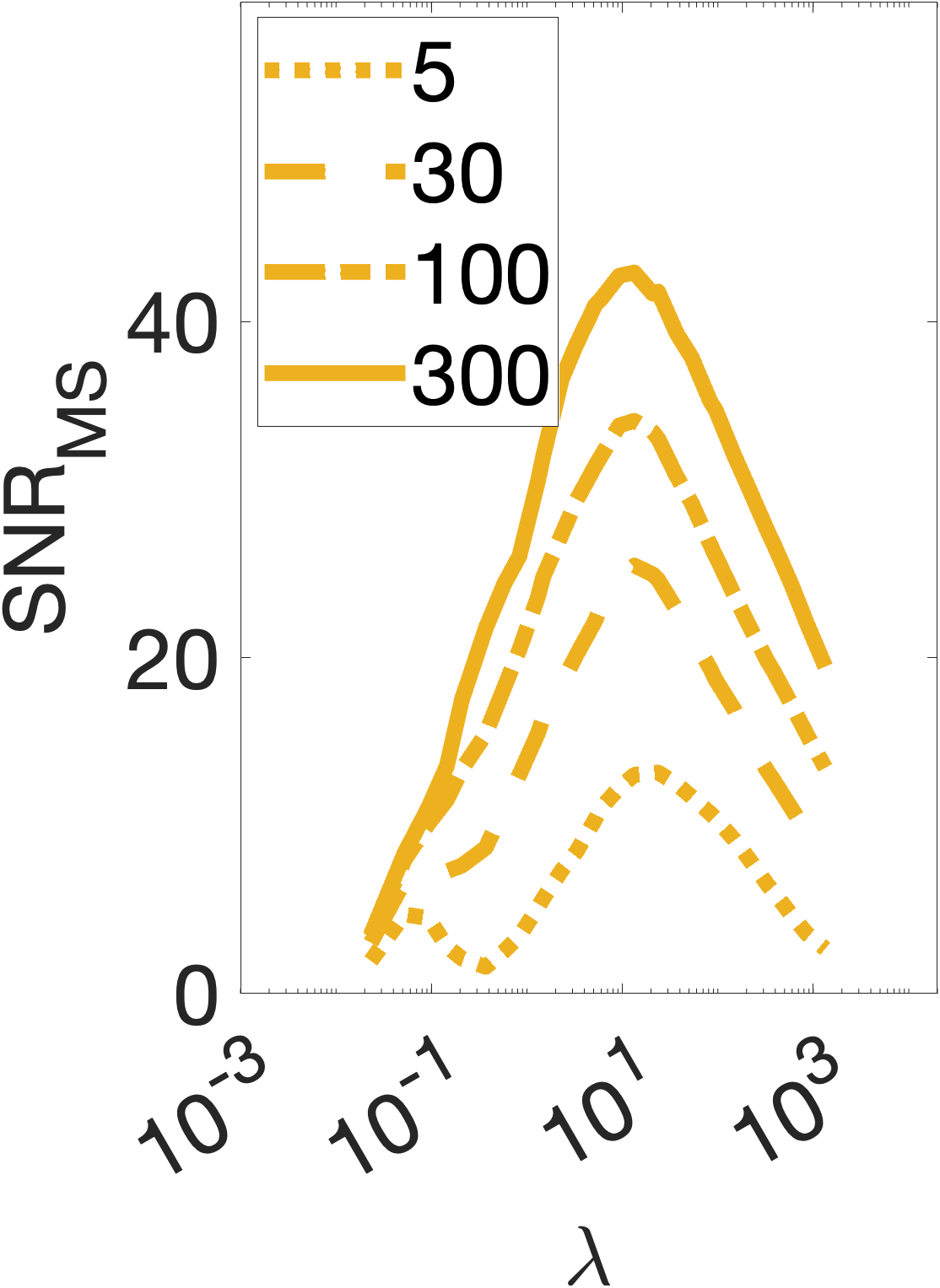}
    \caption{\reviewed{FGLA's performance after various numbers of iterations for $D=8$ and various TF ratios $\lambda$.}}
    \label{fig:convergence_FGLA_lambda}
\end{figure}

To this end, we investigated the interaction between the STFT parameters and the convergence properties of FGLA on the speech dataset. The evaluation considered the Gaussian window, the redundancy at which FGLA performed best, $D=8$, and a wide range of TF ratios, $\lambda \in \{10^{-3}, 10^4\}$. The results were collected after 5, 30, 100, and 300 iterations and are presented in Fig. \ref{fig:convergence_FGLA_lambda}. 

After only five iterations, the range of TF ratios yielding good PR performance can be identified, both in terms of $\mathbf{ODG}$ and, to a lesser extent, of $\mathbf{SNR_{MS}}$. After 30 iterations, this range is clear for both measures. While $\mathbf{SNR_{MS}}$ improved with the increasing number of iteration, $\mathbf{ODG}$ showed ceiling effects after 100 iterations for a wider range of TF ratios. \reviewed{Both measures agreed that PR performance at 100 iterations was very good, even though $\mathbf{SNR_{MS}}$ continued to improve afterwards, at the cost of higher computation time.}

\begin{figure}
    \centering
    \includegraphics[width=0.49\columnwidth]{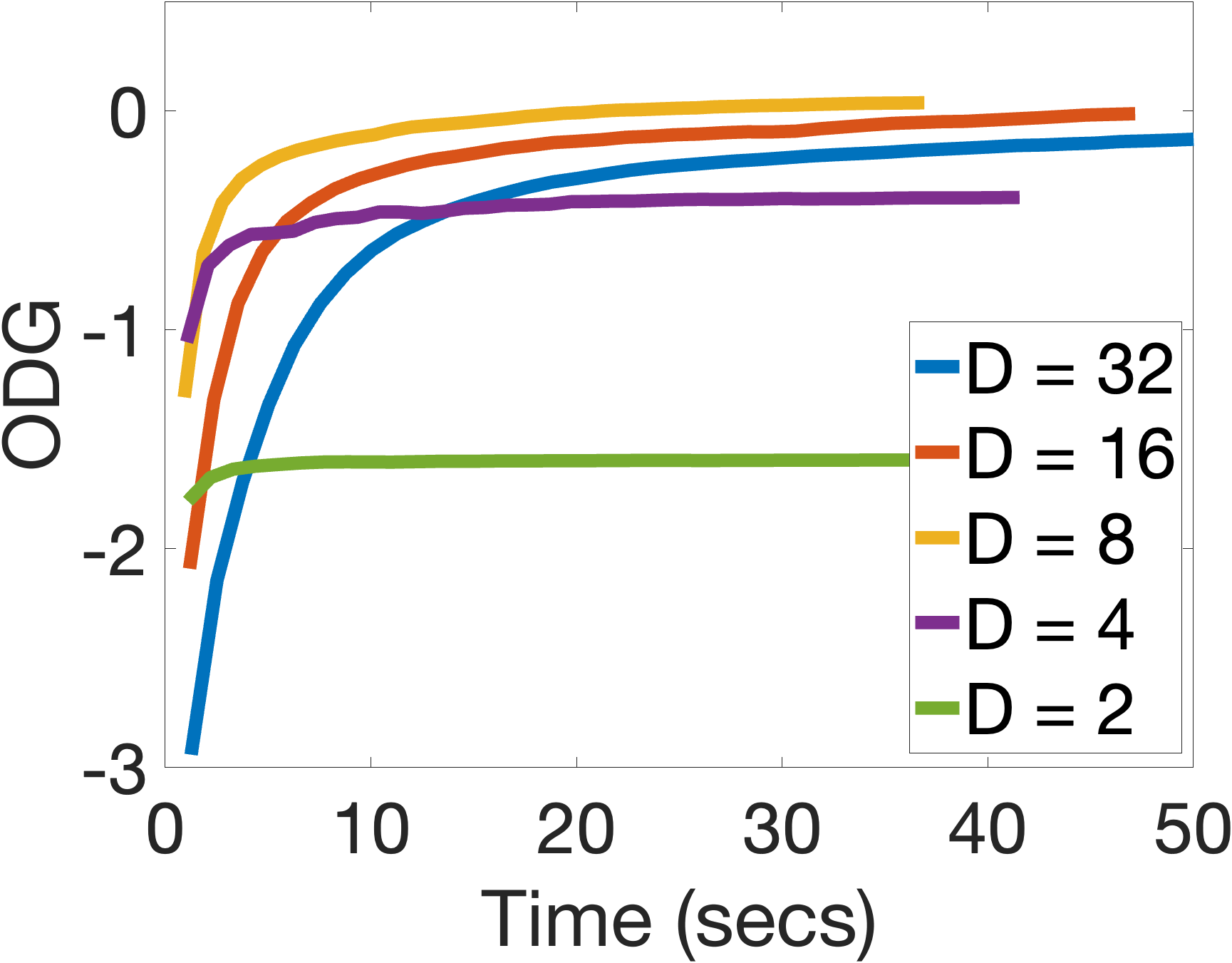}
    \includegraphics[width=0.49\columnwidth]{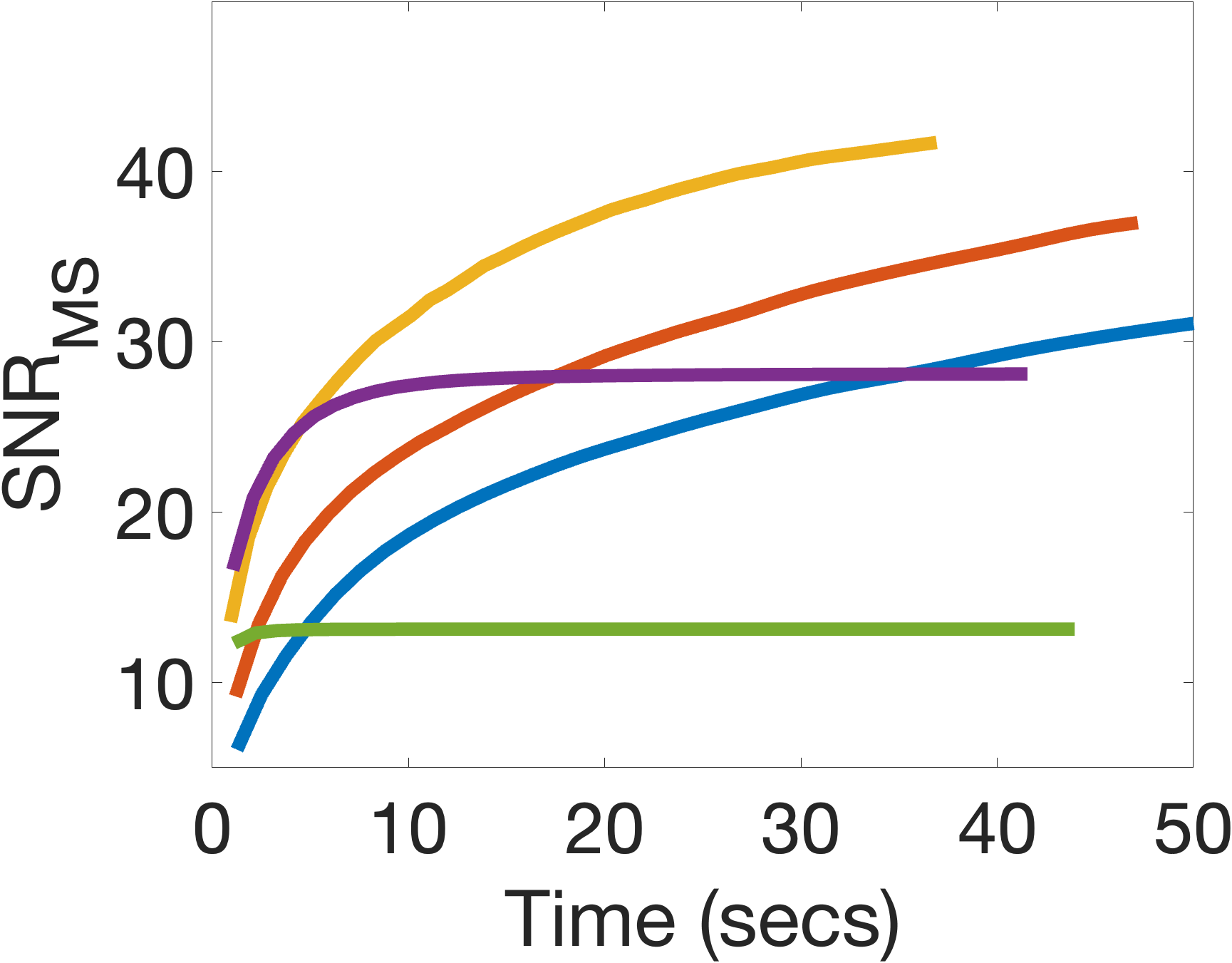}
    \caption{\reviewed{FGLA's performance as a function of the iteration number for five redundancies and $\lambda=3.34$.}}
    \label{fig:timed_convergence_FGLA}
\end{figure}

In the next step, we looked into the time-performance trade-off for a good TF ratio. To this end, we fixed the TF ratio at $\lambda=3.34$ and performed PR with FGLA for redundancies $D \in \{2,4,8,16,32\}$. \reviewed{We set a minimum number of 120 iterations at redundancy $32$, such that we always perform more than the default 100 iterations. Per halving redundancy, we doubled the number of iterations, to have similar computation times per redundancy.} Figure \ref{fig:timed_convergence_FGLA} shows the PR performance plotted against the computational time  consumed in our workstation\footnote{Our workstation is a Windows 10 PC equipped with an Intel i5 7400 processor and 16 GB of RAM. For these experiments we used the MEX backend for LTFAT and the PHASERET toolbox.}. The redundancy $D=8$ resulted in the best time-performance trade-off, with the exception of the first $~5$ seconds at $D=4$, where $\mathbf{SNR_{MS}}$ showed slightly better results.

In conclusion, \reviewed{this experiment reveals} that the optimal range of $\lambda$ for the iterative PR algorithm FGLA can be obtained after as little as 5 iterations, greatly reducing the computation time required to \reviewed{find the optimal $\lambda$ for} a new dataset. We also learned that redundancy 8 does not only perform the best in terms of quality, but it also maximizes the performance/computation time trade-off.

\subsection{\reviewed{Optimal parameters for future applications}}

\reviewed{Our results from the experiments allow us to compose a procedure for finding the optimal set of STFT parameters $(\lambda, D)$ in future applications. This procedure retrieves the range of optimal parameters based on user's input such as the PR algorithm and the audio content. The work flow is as follows:

\begin{enumerate}
    \item Select the phase-retrieval method to evaluate, the audio signals, and the range of $\lambda$ and $D$.
    \item Select error measures to use for evaluation. We recommend a combination of measures such as $\mathbf{SNR_{MS}}$ and $\mathbf{ODG}$.
    \item Compute the STFT magnitudes with a given $(\lambda, D)$, apply the PR method, and compute the average error across all signals.
    \item Increase $D$ and repeat step 3. If the method performs above the given threshold, repeat this step. If it does not, continue with the previous $D$.
    \item Repeat step 3 with smaller and larger $\lambda$ until the performance starts decreasing. In this way, find the range of $\lambda$ for the given $D$ that perform inside of your given threshold.
\end{enumerate}

An implementation of an algorithm following this guidelines is freely available\footnote{\url{https://github.com/andimarafioti/phaseRetrievalEvaluation}}. As a proof of concept, we applied our algorithm to generate parameter sets for some representative use cases and show the obtained parameters in Table \ref{tab:optimal_windows}.}
\begin{table}[h]
    \centering
    \begin{tabular}{|l|l|l|l|l|l|}
         \hline
        Algorithm & Audio & Best for & $D$ & $\lambda$ & $M$ \\
          \hline
        PGHI & Speech & Quality & 32 & 0.14-53.5 & 320-6144\\
        PGHI & Speech & Speed & 16 & 0.65-11.89 & 480-2048\\
        FGLA (300) & Speech & Quality & 8 & 2.32-37.15 & 640-2560\\
        FGLA (50) & Speech & Speed & 8 & 2.32-13.38 & 640-1536\\
        SPSI & Speech & Quality & 8 & 0.83-1.49 & 384-512\\
        SPSI & Speech & Speed & 4 & 1.16-1.67 & 320-384\\
        PGHI & Music & Quality & 32 & 0.21-53.5 & 384-6144\\
        PGHI & Music & Speed & 16 & 1.16-107.0 & 640-6144\\
        FGLA (300) & Music & Quality & 8 &  9.28-334.37 & 1280-7680\\
        FGLA (50) & Music & Speed & 8 & 9.28-95.11 & 1280-4096\\
        SPSI & Music & Quality & 8 & 3.34-20.9 & 768-1920\\
        SPSI & Music & Speed & 4 & 6.68-18.57 & 768-1280 \\
          \hline
    \end{tabular}
    \vspace{1em}
    \caption{Optimal parameter sets for various use cases. The number in parenthesis after FGLA refers to the number of applied iterations. Audio content as in Sec.~\ref{subsec:data}. $M$ was derived from $\lambda$ and the redundancy $D$. The threshold was set to $15$~dB for $\mathbf{SNR_{MS}}$ and to $0.3$ for $\mathbf{ODG}$.}
    \label{tab:optimal_windows}
\end{table}

\section{Conclusions}

We systematically studied the effects of STFT parameters and audio content on the quality of PR algorithms. The goal was to demonstrate the effects, develop guidelines for optimizing STFT parameters, and explain why they improve the PR performance. To this end, we considered three classes of algorithms reconstructing phase from STFT magnitude: iterative (represented by FGLA) and non-iterative with and without a signal model  (represented by SPSI and PGHI, respectively). Our results show that the PR performance depends on the algorithm, the redundancy $D$, the TF ratio $\lambda$, the signal content, and even the window type. Further, we provide guidelines to find the best combinations of the parameters for a specific application. 

As for the algorithms, PGHI showed the best and SPSI the worst performance in terms of SNR and ODG. The performance increased with $D$, with PGHI showing improvements beyond $D$ of 8, and FGLA and SPSI having their performance limited even for $D>8$. We explain this clear advantage of PGHI by its direct exploitation of the theoretical phase-magnitude relation present in the continuous STFT. Still, FGLA has the advantage of providing better results at lower redundancies ($D \leq 8$), however, at the cost of significantly higher computation times. To find the optimal STFT parameters for FGLA, as little as five iterations sufficiently represent the relative performance of the fully converged method. In general, PGHI with $D=32$ seems to be a good choice, achieving a high SNR and an ODG corresponding to the category `imperceptible'. While such high redundancies are currently rarely considered in practice, their advantages on the performance of phase retrieval may push forward their popularity in future systems.

In all three algorithms, the optimal TF ratio was in the range of 0.2 to 20. The particular choice seems to depend on the type of application, though. For example, in our experiment of PR algorithms applied to processed spectrograms, larger $\lambda$ provided an improvement. Also, the audio content requires special consideration as we found higher optimal $\lambda$ required for signals having more energy in the lower frequencies. This can be explained by the need of longer windows for low-frequency content, an issue well-known in the parametrization of STFT for audio applications, which, as our results show, also hold in the problem of phase reconstruction. 

The window type had an effect on the PR performance. While for FGLA and SPSI, that effect was tiny compared to that caused by other parameters, for PGHI, the Gaussian window provided up to 20 dB more SNR compared to other windows. This can be explained by the Gaussian STFT adhering most closely to the PGHI model, which relies on the phase-magnitude relations.

While our results demonstrate how to choose the optimal parameters, many applications receive suboptimal TF representations for PR. Thus, 
future work may consider the development of a system transforming a TF representation computed with a suboptimal set of parameters into TF representations better suited for PR.

\bibliographystyle{IEEEtran}
\bibliography{context}

\begin{thebibliography}{10}
\providecommand{\url}[1]{#1}
\csname url@samestyle\endcsname
\providecommand{\newblock}{\relax}
\providecommand{\bibinfo}[2]{#2}
\providecommand{\BIBentrySTDinterwordspacing}{\spaceskip=0pt\relax}
\providecommand{\BIBentryALTinterwordstretchfactor}{4}
\providecommand{\BIBentryALTinterwordspacing}{\spaceskip=\fontdimen2\font plus
\BIBentryALTinterwordstretchfactor\fontdimen3\font minus
  \fontdimen4\font\relax}
\providecommand{\BIBforeignlanguage}[2]{{%
\expandafter\ifx\csname l@#1\endcsname\relax
\typeout{** WARNING: IEEEtran.bst: No hyphenation pattern has been}%
\typeout{** loaded for the language `#1'. Using the pattern for}%
\typeout{** the default language instead.}%
\else
\language=\csname l@#1\endcsname
\fi
#2}}
\providecommand{\BIBdecl}{\relax}
\BIBdecl

\bibitem{laitinen2013sensitivity}
M.-V. Laitinen, S.~Disch, and V.~Pulkki, ``Sensitivity of human hearing to
  changes in phase spectrum,'' \emph{Journal of the Audio Engineering Society},
  vol.~61, no.~11, pp. 860--877, 2013.

\bibitem{liu1997effects}
L.~Liu, J.~He, and G.~Palm, ``Effects of phase on the perception of
  intervocalic stop consonants,'' \emph{Speech Communication}, vol.~22, no.~4,
  pp. 403--417, 1997.

\bibitem{paliwal2005usefulness}
K.~K. Paliwal and L.~D. Alsteris, ``On the usefulness of {STFT} phase spectrum
  in human listening tests,'' \emph{Speech Communication}, vol.~45, no.~2, pp.
  153--170, 2005.

\bibitem{oppenheim1981importance}
A.~V. Oppenheim and J.~S. Lim, ``The importance of phase in signals,''
  \emph{Proceedings of the IEEE}, vol.~69, no.~5, pp. 529--541, 1981.

\bibitem{Allen1977}
J.~Allen, ``Short term spectral analysis, synthesis, and modification by
  discrete {F}ourier transform,'' \emph{IEEE Transactions on Acoustics, Speech,
  and Signal Processing}, vol.~25, no.~3, pp. 235--238, 1977.

\bibitem{rawe90}
J.~Wexler and S.~Raz, ``Discrete {G}abor expansions,'' \emph{Signal
  Processing}, vol.~21, no.~3, pp. 207 -- 220, 1990.

\bibitem{vincent2018audio}
E.~Vincent, T.~Virtanen, and S.~Gannot, \emph{Audio source separation and
  speech enhancement}.\hskip 1em plus 0.5em minus 0.4em\relax John Wiley \&
  Sons, 2018.

\bibitem{Chowdhury2019}
S.~Chowdhury, A.~V. Portabella, V.~Haunschmid, and G.~Widmer, ``Towards
  explainable music emotion recognition: The route via mid-level features,'' in
  \emph{Proc. of the 20th International Society for Music Information Retrieval
  Conference}, 2019, pp. 237--243.

\bibitem{pepino2018sourcesep}
L.~Pepino and L.~Bender, ``Separación de fuentes musicales mediante redes
  neuronales convolucionales con múltiples decodificadores,'' in
  \emph{Jornadas de Audio, Acústica y Sonido}.\hskip 1em plus 0.5em minus
  0.4em\relax UNTREF, 2018.

\bibitem{ghose2020Enabling}
S.~{Ghose} and J.~J. {Prevost}, ``Enabling an {IoT} system of systems through
  auto sound synthesis in silent video with {DNN},'' in \emph{2020 IEEE 15th
  International Conference of System of Systems Engineering (SoSE)}, 2020, pp.
  563--568.

\bibitem{magron2018reducing}
P.~Magron, K.~Drossos, S.~Mimilakis, and T.~Virtanen, ``Reducing interference
  with phase recovery in {DNN}-based monaural singing voice separation,'' in
  \emph{Proc. of INTERSPEECH}, 2018.

\bibitem{Liu2020Unified}
B.~{Liu}, A.~{Cao}, and H.~{Kim}, ``Unified signal compression using generative
  adversarial networks,'' in \emph{ICASSP 2020 - 2020 IEEE International
  Conference on Acoustics, Speech and Signal Processing (ICASSP)}, 2020, pp.
  3177--3181.

\bibitem{marafioti2020gacela}
A.~Marafioti, P.~Majdak, N.~Holighaus, and N.~Perraudin, ``{GACELA} -- {A}
  generative adversarial context encoder for long audio inpainting,''
  \emph{IEEE Journal of Selected Topics in Signal Processing, Special issue on
  reconstruction of audio from incomplete or highly degraded observations}, p.
  to appear, 2020.

\bibitem{engel2019gansynth}
J.~Engel, K.~K. Agrawal, S.~Chen, I.~Gulrajani, C.~Donahue, and A.~Roberts,
  ``{GANSynth}: Adversarial neural audio synthesis,'' in \emph{Proc. of ICLR},
  2019.

\bibitem{gerkmann2015phase}
T.~Gerkmann, M.~Krawczyk-Becker, and J.~Le~Roux, ``Phase processing for
  single-channel speech enhancement: History and recent advances,'' \emph{IEEE
  signal processing Magazine}, vol.~32, no.~2, pp. 55--66, 2015.

\bibitem{mowlaee2017single}
P.~Mowlaee, J.~Kulmer, J.~Stahl, and F.~Mayer, \emph{Single Channel Phase-Aware
  Signal Processing in Speech Communication: Theory and Practice}.\hskip 1em
  plus 0.5em minus 0.4em\relax {Wiley-IEEE Press}, 2017.

\bibitem{magron2015phaserecovery}
P.~{Magron}, R.~{Badeau}, and B.~{David}, ``Phase recovery in {NMF} for audio
  source separation: An insightful benchmark,'' in \emph{2015 IEEE
  International Conference on Acoustics, Speech and Signal Processing
  (ICASSP)}, 2015, pp. 81--85.

\bibitem{marafioti2019context}
A.~Marafioti, N.~Perraudin, N.~Holighaus, and P.~Majdak, ``A context encoder
  for audio inpainting,'' \emph{IEEE/ACM Transactions on Audio, Speech, and
  Language Processing}, vol.~27, no.~12, pp. 2362--2372, 2019.

\bibitem{marafioti2019audio}
A.~Marafioti, N.~Holighaus, P.~Majdak, and N.~Perraudin, ``Audio inpainting of
  music by means of neural networks,'' in \emph{Audio Engineering Society
  Convention 146}, Mar 2019.

\bibitem{harrison1993phase}
R.~W. Harrison, ``Phase problem in crystallography,'' \emph{Journal of the
  Optical Society of America A}, vol.~10, no.~5, pp. 1046--1055, 1993.

\bibitem{miao2008extending}
J.~Miao, T.~Ishikawa, Q.~Shen, and T.~Earnest, ``Extending {X}-ray
  crystallography to allow the imaging of noncrystalline materials, cells, and
  single protein complexes,'' \emph{Annual Review of Physical Chemistry},
  vol.~59, pp. 387--410, 2008.

\bibitem{fogel2016phase}
F.~Fogel, I.~Waldspurger, and A.~d’Aspremont, ``Phase retrieval for imaging
  problems,'' \emph{Mathematical programming computation}, vol.~8, no.~3, pp.
  311--335, 2016.

\bibitem{shechtman2015phase}
Y.~Shechtman, Y.~C. Eldar, O.~Cohen, H.~N. Chapman, J.~Miao, and M.~Segev,
  ``Phase retrieval with application to optical imaging: a contemporary
  overview,'' \emph{IEEE signal processing magazine}, vol.~32, no.~3, pp.
  87--109, 2015.

\bibitem{Bendory2017}
T.~Bendory, R.~Beinert, and Y.~C. Eldar, \emph{Fourier Phase Retrieval:
  Uniqueness and Algorithms}.\hskip 1em plus 0.5em minus 0.4em\relax Cham:
  Springer International Publishing, 2017, pp. 55--91.

\bibitem{balan2006signal}
R.~Balan, P.~Casazza, and D.~Edidin, ``On signal reconstruction without
  phase,'' \emph{Applied and Computational Harmonic Analysis}, vol.~20, no.~3,
  pp. 345--356, 2006.

\bibitem{nawab1983signal}
S.~Nawab, T.~Quatieri, and J.~Lim, ``Signal reconstruction from short-time
  {F}ourier transform magnitude,'' \emph{IEEE Transactions on Acoustics,
  Speech, and Signal Processing}, vol.~31, no.~4, pp. 986--998, 1983.

\bibitem{bojarovska2016phase}
I.~Bojarovska and A.~Flinth, ``Phase retrieval from {G}abor measurements,''
  \emph{Journal of {F}ourier Analysis and Applications}, vol.~22, no.~3, pp.
  542--567, 2016.

\bibitem{jaganathan2016stft}
K.~Jaganathan, Y.~C. Eldar, and B.~Hassibi, ``{STFT} phase retrieval:
  Uniqueness guarantees and recovery algorithms,'' \emph{IEEE Journal of
  selected topics in signal processing}, vol.~10, no.~4, pp. 770--781, 2016.

\bibitem{li2017phase}
L.~Li, C.~Cheng, D.~Han, Q.~Sun, and G.~Shi, ``Phase retrieval from
  multiple-window short-time {F}ourier measurements,'' \emph{IEEE Signal
  Processing Letters}, vol.~24, no.~4, pp. 372--376, 2017.

\bibitem{alaifari2019ill}
R.~Alaifari and M.~Wellershoff, ``Ill-conditionedness of discrete {G}abor phase
  retrieval and a possible remedy,'' in \emph{2019 13th International
  conference on Sampling Theory and Applications (SampTA)}, 2019, pp. 1--4.

\bibitem{waldspurger2018phase}
I.~Waldspurger, ``Phase retrieval with random gaussian sensing vectors by
  alternating projections,'' \emph{IEEE Transactions on Information Theory},
  vol.~64, no.~5, pp. 3301--3312, 2018.

\bibitem{griffin1984signal}
D.~Griffin and J.~Lim, ``Signal estimation from modified short-time {F}ourier
  transform,'' \emph{IEEE Transactions on Acoustics, Speech and Signal
  Processing}, vol.~32, no.~2, pp. 236--243, 1984.

\bibitem{masuyama2020deepGriffin}
Y.~{Masuyama}, K.~{Yatabe}, Y.~{Koizumi}, Y.~{Oikawa}, and N.~{Harada}, ``Deep
  {G}riffin–{L}im iteration: Trainable iterative phase reconstruction using
  neural network,'' \emph{IEEE Journal of Selected Topics in Signal
  Processing}, pp. 1--1, 2020.

\bibitem{masuyama2019deep}
Y.~Masuyama, K.~Yatabe, Y.~Koizumi, Y.~Oikawa, and N.~Harada, ``Deep
  {G}riffin-{L}im iteration,'' in \emph{Proc. of ICASSP}.\hskip 1em plus 0.5em
  minus 0.4em\relax IEEE, 2019, pp. 61--65.

\bibitem{Masuyama2020ICASSP}
Y.~{Masuyama}, K.~{Yatabe}, Y.~{Koizumi}, Y.~{Oikawa}, and N.~{Harada}, ``Phase
  reconstruction based on recurrent phase unwrapping with deep neural
  networks,'' in \emph{ICASSP 2020 - 2020 IEEE International Conference on
  Acoustics, Speech and Signal Processing (ICASSP)}, 2020, pp. 826--830.

\bibitem{Masuyama2019griffinLim}
Y.~{Masuyama}, K.~{Yatabe}, and Y.~{Oikawa}, ``{G}riffin–{L}im like phase
  recovery via alternating direction method of multipliers,'' \emph{IEEE Signal
  Processing Letters}, vol.~26, no.~1, pp. 184--188, 2019.

\bibitem{le2010fast}
J.~Le~Roux, H.~Kameoka, N.~Ono, and S.~Sagayama, ``Fast signal reconstruction
  from magnitude {STFT} spectrogram based on spectrogram consistency,'' in
  \emph{Proc. Int. Conf. Digital Audio Effects}, vol.~10, 2010.

\bibitem{Perraudin2013griffin}
N.~Perraudin, P.~Balazs, and P.~L. S{\o}ndergaard, ``A fast {G}riffin-{L}im
  algorithm,'' in \emph{Applications of Signal Processing to Audio and
  Acoustics (WASPAA), 2013 IEEE Workshop on}.\hskip 1em plus 0.5em minus
  0.4em\relax IEEE, 2013, pp. 1--4.

\bibitem{Zhu2006}
X.~{Zhu}, G.~T. {Beauregard}, and L.~{Wyse}, ``Real-time iterative spectrum
  inversion with look-ahead,'' in \emph{2006 IEEE International Conference on
  Multimedia and Expo}, 2006, pp. 229--232.

\bibitem{mimilakis2018monaural}
S.~I. Mimilakis, K.~Drossos, J.~F. Santos, G.~Schuller, T.~Virtanen, and
  Y.~Bengio, ``Monaural singing voice separation with skip-filtering
  connections and recurrent inference of time-frequency mask,'' in \emph{2018
  IEEE International Conference on Acoustics, Speech and Signal Processing
  (ICASSP)}, 2018, pp. 721--725.

\bibitem{vasquez2019melnet}
S.~Vasquez and M.~Lewis, ``Mel{N}et: A generative model for audio in the
  frequency domain,'' in \emph{Proc. of ICLR}, 2020.

\bibitem{beauregard2015single}
G.~T. Beauregard, M.~Harish, and L.~Wyse, ``Single pass spectrogram
  inversion,'' in \emph{2015 IEEE international conference on digital signal
  processing (DSP)}, 2015, pp. 427--431.

\bibitem{magron2015phase}
P.~Magron, R.~Badeau, and B.~David, ``Phase reconstruction of spectrograms with
  linear unwrapping: application to audio signal restoration,'' in \emph{2015
  23rd European Signal Processing Conference (EUSIPCO)}, 2015, pp. 1--5.

\bibitem{ltfatnote040}
Z.~Pr\r{u}\v{s}a, P.~Balazs, and P.~L. S{\o}ndergaard, ``A noniterative method
  for reconstruction of phase from {STFT} magnitude,'' \emph{IEEE/ACM Trans. on
  Audio, Speech, and Lang. Process.}, vol.~25, no.~5, May 2017.

\bibitem{laroche1999improved}
J.~Laroche and M.~Dolson, ``Improved phase vocoder time-scale modification of
  audio,'' \emph{IEEE Transactions on Speech and Audio processing}, vol.~7,
  no.~3, pp. 323--332, 1999.

\bibitem{portnoff1979magnitude}
M.~Portnoff, ``Magnitude-phase relationships for short-time {F}ourier
  transforms based on {G}aussian analysis windows,'' in \emph{ICASSP'79. IEEE
  International Conference on Acoustics, Speech, and Signal Processing},
  vol.~4.\hskip 1em plus 0.5em minus 0.4em\relax IEEE, 1979, pp. 186--189.

\bibitem{marafioti2019adversarial}
A.~Marafioti, N.~Perraudin, N.~Holighaus, and P.~Majdak, ``Adversarial
  generation of time-frequency features with application in audio synthesis,''
  in \emph{Proc. of the 36th ICML}, K.~Chaudhuri and R.~Salakhutdinov, Eds.,
  vol.~97.\hskip 1em plus 0.5em minus 0.4em\relax Long Beach, California, USA:
  PMLR, 09--15 Jun 2019, pp. 4352--4362.

\bibitem{mowlaee2016advances}
P.~Mowlaee, R.~Saeidi, and Y.~Stylianou, ``Advances in phase-aware signal
  processing in speech communication,'' \emph{Speech Communication}, vol.~81,
  pp. 1--29, 2016.

\bibitem{ltfatnote043}
Z.~Pr\r{u}\v{s}a and P.~L. S{\o}ndergaard, ``Real-time spectrogram inversion
  using phase gradient heap integration,'' in \emph{Proc. Int. Conf. Digital
  Audio Effects (DAFx-16)}, Sep 2016, pp. 17--21.

\bibitem{holighaus2019char}
N.~{Holighaus}, G.~{Koliander}, Z.~{Pr{\r{u}}{\v{s}}a}, and L.~D. {Abreu},
  ``Characterization of analytic wavelet transforms and a new phaseless
  reconstruction algorithm,'' \emph{IEEE Transactions on Signal Processing},
  vol.~67, no.~15, pp. 3894--3908, 2019.

\bibitem{holighaus2019non}
N.~Holighaus, G.~Koliander, L.~D. Abreu, and Z.~Pru{\v{s}}a, ``Non-iterative
  phaseless reconstruction from wavelet transform magnitude,'' in
  \emph{Proceedings of the 22nd International Conference on Digital Audio
  Effects, Birmingham, UK}, 2019, pp. 2--6.

\bibitem{po76}
M.~{P}ortnoff, ``{I}mplementation of the digital phase vocoder using the fast
  {F}ourier transform,'' \emph{{I}{E}{E}{E} {T}rans. {A}coust. {S}peech
  {S}ignal {P}rocess.}, vol.~24, no.~3, pp. 243--248, 1976.

\bibitem{auger2012phase}
F.~Auger, {\'E}.~Chassande-Mottin, and P.~Flandrin, ``On phase-magnitude
  relationships in the short-time {F}ourier transform.'' \emph{IEEE Signal
  Process. Lett.}, vol.~19, no.~5, pp. 267--270, 2012.

\bibitem{gr01}
K.~{G}r{\"o}chenig, \emph{{F}oundations of {T}ime-{F}requency {A}nalysis}, ser.
  {A}ppl. {N}umer. {H}armon. {A}nal.\hskip 1em plus 0.5em minus 0.4em\relax
  {B}irkh{\"a}user, 2001.

\bibitem{2020SciPy-NMeth}
P.~{Virtanen}, R.~{Gommers}, T.~E. {Oliphant}, M.~{Haberland}, T.~{Reddy},
  D.~{Cournapeau}, E.~{Burovski}, P.~{Peterson}, W.~{Weckesser}, J.~{Bright},
  S.~J. {van der Walt}, M.~{Brett}, J.~{Wilson}, K.~{Jarrod Millman},
  N.~{Mayorov}, A.~R.~J. {Nelson}, E.~{Jones}, R.~{Kern}, E.~{Larson},
  C.~{Carey}, {\.I}.~{Polat}, Y.~{Feng}, E.~W. {Moore}, J.~{Vand erPlas},
  D.~{Laxalde}, J.~{Perktold}, R.~{Cimrman}, I.~{Henriksen}, E.~A. {Quintero},
  C.~R. {Harris}, A.~M. {Archibald}, A.~H. {Ribeiro}, F.~{Pedregosa}, P.~{van
  Mulbregt}, and {Contributors}, ``{SciPy 1.0: Fundamental Algorithms for
  Scientific Computing in Python},'' \emph{Nature Methods}, vol.~17, pp.
  261--272, 2020.

\bibitem{ltfatnote030}
Z.~Pr\r{u}\v{s}a, P.~L. S{\o}ndergaard, N.~Holighaus, C.~Wiesmeyr, and
  P.~Balazs, ``The large time-frequency analysis toolbox 2.0,'' in \emph{Sound,
  Music, and Motion}, ser. LNCS.\hskip 1em plus 0.5em minus 0.4em\relax
  Springer International Publishing, 2014, pp. 419--442.

\bibitem{pytorch}
A.~Paszke, S.~Gross, F.~Massa, A.~Lerer, J.~Bradbury, G.~Chanan, T.~Killeen,
  Z.~Lin, N.~Gimelshein, L.~Antiga, A.~Desmaison, A.~Kopf, E.~Yang, Z.~DeVito,
  M.~Raison, A.~Tejani, S.~Chilamkurthy, B.~Steiner, L.~Fang, J.~Bai, and
  S.~Chintala, ``Pytorch: An imperative style, high-performance deep learning
  library,'' in \emph{Advances in Neural Information Processing Systems
  32}.\hskip 1em plus 0.5em minus 0.4em\relax Curran Associates, Inc., 2019,
  pp. 8024--8035.

\bibitem{best03}
T.~{S}trohmer and S.~{B}eaver, ``{O}ptimal {O}{F}{D}{M} system design for
  time-frequency dispersive channels,'' \emph{{I}{E}{E}{E} {T}rans. {C}omm.},
  vol.~51, no.~7, pp. 1111--1122, {J}uly 2003.

\bibitem{faulhuber2017optimal}
M.~Faulhuber and S.~Steinerberger, ``Optimal {G}abor frame bounds for separable
  lattices and estimates for {J}acobi theta functions,'' \emph{Journal of
  Mathematical Analysis and Applications}, vol. 445, no.~1, pp. 407--422, 2017.

\bibitem{st98-1}
T.~{S}trohmer, ``{N}umerical algorithms for discrete {G}abor expansions,'' in
  \emph{{G}abor Analysis and Algorithms: Theory and Applications}, ser. Applied
  and Numerical Harmonic Analysis, H.~G. {F}eichtinger and T.~{S}trohmer,
  Eds.\hskip 1em plus 0.5em minus 0.4em\relax {B}irkh{\"a}user {B}oston, 1998,
  pp. 267--294.

\bibitem{janssen1997continuous}
A.~Janssen, ``From continuous to discrete {W}eyl-{H}eisenberg frames through
  sampling,'' \emph{Journal of {F}ourier Analysis and Applications}, vol.~3,
  no.~5, pp. 583--596, 1997.

\bibitem{ljspeech17}
K.~Ito and L.~Johnson, ``The {LJ} speech dataset,''
  \url{https://keithito.com/LJ-Speech-Dataset/}, 2017.

\bibitem{raffelthesis}
C.~Raffel, ``Learning-based methods for comparing sequences, with applications
  to audio-to-{MIDI} alignment and matching,'' Ph.D. dissertation, 2016.

\bibitem{prettymidi}
C.~Raffel and D.~P.~W. Ellis, ``Intuitive analysis, creation and manipulation
  of {MIDI} data with pretty\_midi,'' in \emph{Proc. of the 15th ISMIR, Late
  Breaking and Demo Papers}, 2014.

\bibitem{fma2017}
M.~Defferrard, K.~Benzi, P.~Vandergheynst, and X.~Bresson, ``Fma: A dataset for
  music analysis,'' in \emph{18th International Society for Music Information
  Retrieval Conference}, 2017.

\bibitem{ltfatnote045}
Z.~Pr\r{u}\v{s}a, ``The phase retrieval toolbox,'' in \emph{{AES} International
  Conference On Semantic Audio}, Erlangen, Germany, June 2017.

\bibitem{gerchberg1972practical}
R.~W. Gerchberg and W.~O. Saxton, ``A practical algorithm for the determination
  of phase from image and diffraction plane pictures,'' \emph{Optik}, vol.~35,
  pp. 237--246, 1972.

\bibitem{thiede2000peaq}
T.~Thiede, W.~C. Treurniet, R.~Bitto, C.~Schmidmer, T.~Sporer, J.~G. Beerends,
  and C.~Colomes, ``{PEAQ}-{T}he {ITU} standard for objective measurement of
  perceived audio quality,'' \emph{J. Aud. Eng. Soc.}, vol.~48, no. 1/2, pp.
  3--29, 2000.

\bibitem{recommendation20011387}
{ITU-R Recommendation}, ``1387: Method for objective measurements of perceived
  audio quality,'' \emph{International Telecommunication Union, Geneva,
  Switzerland}, 2001.

\bibitem{kabal2002examination}
P.~Kabal \emph{et~al.}, ``An examination and interpretation of {ITU-R BS}.
  1387: Perceptual evaluation of audio quality,'' \emph{TSP Lab Technical
  Report, Dept. Electrical \& Computer Engineering, McGill University}, pp.
  1--89, 2002.

\bibitem{huber2006pemo}
R.~Huber and B.~Kollmeier, ``Pemo-q—a new method for objective audio quality
  assessment using a model of auditory perception,'' \emph{IEEE Transactions on
  audio, speech, and language processing}, vol.~14, no.~6, pp. 1902--1911,
  2006.

\bibitem{emiya2011subjective}
V.~Emiya, E.~Vincent, N.~Harlander, and V.~Hohmann, ``Subjective and objective
  quality assessment of audio source separation,'' \emph{IEEE Transactions on
  Audio, Speech, and Language Processing}, vol.~19, no.~7, pp. 2046--2057,
  2011.

\end{thebibliography}

\end{document}